\documentclass[12pt,fleqn]{article}

\DeclareMathAlphabet{\mathpzc}{OT1}{pzc}{m}{it}
\usepackage{times}
\usepackage{txfonts}
\usepackage{amssymb}
\usepackage{setspace}
\usepackage{graphicx}
\usepackage{harvard}
\renewcommand{\harvardand}{, and}
\citationmode{abbr}
\usepackage[usenames,dvipsnames]{color}

\begin{document}
\date{}
\title{Chapter 7: Computational design of chemical nanosensors: Transition metal doped single-walled carbon nanotubes}
\author{\normalsize{Duncan J. Mowbray\thanks{Nano-Bio Spectroscopy Group and ETSF Scientific Development Centre, Depto.~F{\'{\i}}sica de Materiales, Universidad del Pa{\'{\i}}s Vasco and DIPC, Avenida de Tolosa 72, E-20018 San Sebasti{\'{a}}n, Spain; Phone:+34 943 01 8288 Fax:+34 943 01 8390 E-mail: duncan.mowbray@gmail.com}, 
Juan Mar{\'{\i}}a Garc{\'{\i}}a-Lastra\thanks{Nano-Bio Spectroscopy Group and ETSF Scientific Development Centre, Depto.~F{\'{\i}}sica de Materiales, Universidad del Pa{\'{\i}}s Vasco, Centro de F{\'{\i}}sica de Materiales CSIC-UPV/EHU- MPC and DIPC, Av. Tolosa 72, E-20018 San Sebasti{\'{a}}n, Spain; Center for Atomic-scale Materials Design, Department of Physics,  Technical University of Denmark, DK-2800 Kgs.~Lyngby, Denmark; Phone:+45 4525 3209 Fax:+45 4593 2399 E-mail: jumagala@fysik.dtu.dk}, Iker Larraza Arocena\thanks{Nano-Bio Spectroscopy Group and ETSF Scientific Development Centre, Depto.~F{\'{\i}}sica de Materiales, Universidad del Pa{\'{\i}}s Vasco and DIPC, Avenida de Tolosa 72, E-20018 San Sebasti{\'{a}}n, Spain; E-mail: ilarrazaarocena@gmail.com},}\\
\normalsize{\'{A}ngel Rubio\thanks{Nano-Bio Spectroscopy Group and ETSF Scientific Development Centre, Depto.~F{\'{\i}}sica de Materiales, Universidad del Pa{\'{\i}}s Vasco, Centro de F{\'{\i}}sica de Materiales CSIC-UPV/EHU- MPC and DIPC, Avenida de Tolosa 72, E-20018 San Sebasti{\'{a}}n, Spain; 
Phone:+34 943 01 8292 Fax:+34 943 01 8390 E-mail: angel.rubio@ehu.es}, Kristian S. Thygesen\thanks{Center for Atomic-scale Materials Design, Department of Physics,  Technical University of Denmark, DK-2800 Kgs.~Lyngby, Denmark; Phone:+45 4525 3188 Fax:+45 4593 2399 E-mail: thygesen@fysik.dtu.dk}, 
Karsten W. Jacobsen
\thanks{Center for Atomic-scale Materials Design, Department of Physics,  Technical University of Denmark, DK-2800 Kgs.~Lyngby, Denmark; Phone:+45 4525 3186 Fax:+45 4593 2399 E-mail: kwj@fysik.dtu.dk}}}
\maketitle

\clearpage
\newpage
\section*{Summary}
We present a general approach to the computational design of
nanostructured chemical sensors. The scheme is based on identification
and calculation of microscopic descriptors (design parameters) which
are used as input to a thermodynamic model to obtain the relevant
macroscopic properties. In particular, we consider the
functionalization of a (6,6) metallic armchair single-walled carbon nanotube (SWNT) by nine different $3d$
transition metal (TM) atoms occupying three types of vacancies. For
six gas molecules (N$_{{2}}$, O$_{{2}}$, H$_{{2}}$O,
CO, NH$_{{3}}$, H$_{{2}}$S) we calculate the binding  
energy and change in conductance due to adsorption on each of the 27
TM sites. For a given type of TM functionalization, this allows us to
obtain the equilibrium coverage and change in conductance as a function
of the partial pressure of the ``target'' molecule in a background of
atmospheric air.  Specifically, we show how Ni and Cu doped metallic (6,6) SWNTs may work as effective multifunctional sensors for both CO and NH$_3$.  In this way, the scheme presented allows one to obtain macroscopic device characteristics and performance data for nanoscale (in this case SWNT) based devices. 


%
\clearpage
\newpage
\tableofcontents
\clearpage
\newpage
\section{Introduction}

Detecting specific chemical species at small concentrations is of fundamental importance for many industrial and scientific processes, medical applications, and environmental monitoring.  Nanostructured materials are ideally suited for sensor applications because of their large surface to volume ratio, making them sensitive to the adsorption of individual molecules.  

\begin{figure}[!thb]
\includegraphics[width=\columnwidth]{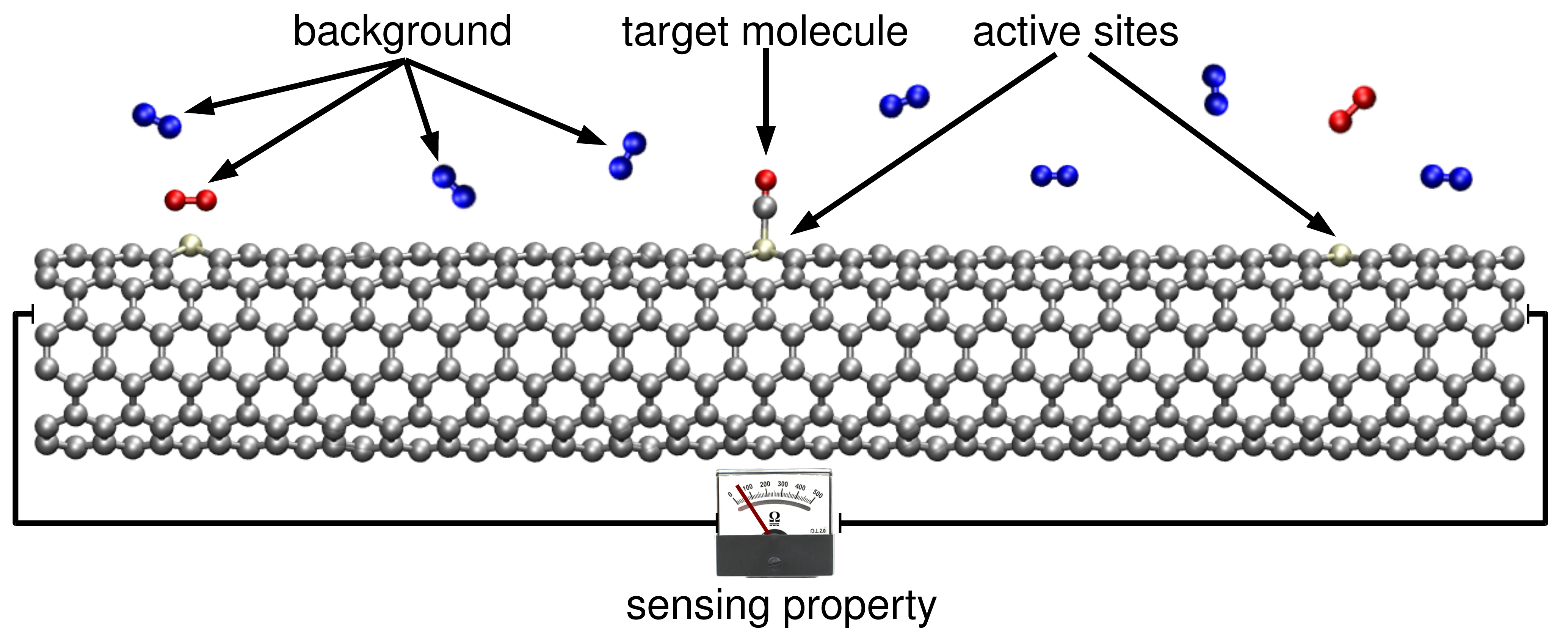}
\caption{Schematic of a chemical sensor consisting of active sites (metal dopants in a (6,6) SWNT), a target molecule (CO), a background (atmospheric air), and a sensing property (resistance) (Mowbray et~al.~2010)
.}
\label{Fig1}
\end{figure}

At a general level, any nanosensing system consists of the following
four main components: 
\begin{enumerate}
\item a ``target molecule'' to be detected,
\item an ``active site'' where the target molecule may adsorb on the sensor,
\item a ``sensing property'' which changes upon adsorption of the
target molecule,
\item a ``background'' of adsorbing molecules which
make up the background signal. 
\end{enumerate}
The active site must be designed so
that adsorption of the target molecule in the presence of the
background is sufficient to change the sensing property.  These four main components of a chemical sensor are shown schematically in Figure \ref{Fig1}, for the case of a transition metal (TM) doped (6,6) metallic armchair single-walled carbon nanotube (SWNT) measuring CO under atmospheric conditions.

Performing a screening study of possible SWNT dopings experimentally would be quite
demanding, especially considering the great variety of potential
active sites and target molecules one may wish to detect.   On the
other hand, computational screening studies based on density
functional theory (DFT) calculations have recently been shown to be
quite effective at shortening lists of potential candidates, which may
then be studied experimentally.   

The energetics obtained from DFT calculations using a generalized gradient approximation (GGA) for the exchange and correlation (xc)-functional
are sufficiently accurate to provide a quantitative description of the
stability of an active site, and the adsorption energies for both the
target molecule and the background.  From this data we may use kinetic 
modeling to estimate the equilibrium coverage of active sites by target molecules as a function of their pressure in the presence of the background.
This allows us to quickly screen a wide variety of active sites, and
rule out those which are completely oxidized in air, or too unstable
at normal temperatures.   

At the same time, changes in a nanosensor's resistance are often used to detect
adsorption of a target molecule.  Trends in the resistance may be
reasonably described using the non-equilibrium Green's function (NEGF)
methodology.  In this way we may estimate the resistance per active
site in the background, and hence determine the change in resistance
of the active site as a function of target molecule concentration.

Using resistance as a sensing property has two main advantages.
First, it is a non-intrusive measure, which should not significantly
influence the adsorption of the target molecules.  Second, small
changes in the coverage of target molecules ($\sim 0.1\%$) often yield
large changes in resistance ($\sim 1 \Omega$/site).   

\begin{figure}[!h]
\centering
\includegraphics[scale=0.5]{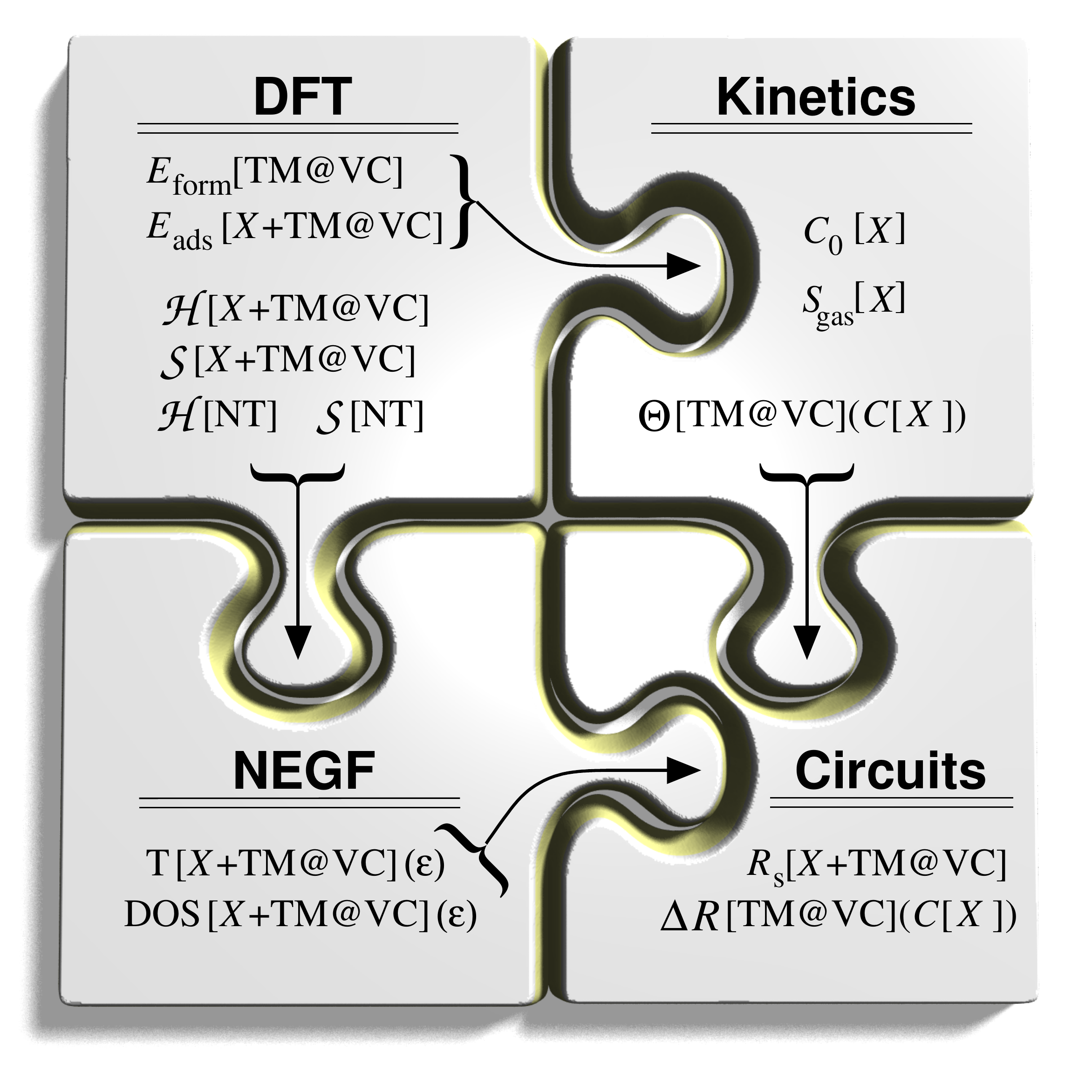}
\caption{Schematic of the overall procedure for modeling a nanosensor, fitting together like four pieces of a puzzle, density functional theory (DFT) calculations, kinetic modeling, non-equilibrium Green's function (NEGF) calculations, and circuit theory, as described in the text.}\label{Method}
\end{figure}

Overall, this procedure fits together like four ``puzzle pieces'',  DFT calculations, kinetic modeling, NEGF results and circuit theory, to calculate the average change in resistance $\Delta R$ as a function of a target molecule's concentration $C[X]$.  This methodology for modeling the chemical sensor is depicted schematically in Figure \ref{Method}.    

From DFT we obtain the formation energies $E_{\mathrm{form}}$ of an active site (a 3$d$ TM occupied vacancy (TM@VC)), the adsorption energies $E_{\mathrm{ads}}$ for species $X$ on the active site ($X$+TM@VC), and the Hamiltonian $\mathpzc{H}$ and overlap matrices $\mathpzc{S}$ for the scattering region with $X$ adsorbed ($X$+TM@VC) and the leads (a pristine (6,6) SWNT (NT)).  The energetics from DFT are then inputted into a kinetic model for the active site in thermodynamic equilibrium with concentrations $C_0$, and gas phase entropies $S_{\mathrm{gas}}$,  for each species taken from experiment \cite{CRCHandbook}.  From this we obtain the fractional coverage $\Theta$ per active site (TM@VC) as a function of the target molecule's concentration $C[X]$.  

At the same time, the Hamiltonian $\mathpzc{H}$ and overlap $\mathpzc{S}$ matrices obtained from DFT are used within the NEGF methodology to obtain the density of states (DOS) and transmission probability T for each species $X$ on an active site ($X$+TM@VC), as a function of energy $\varepsilon$.  Finally the coverage and transmission probabilities are used within simple circuit theory to estimate the scattering resistance $R_s$ for species $X$ on an active site ($X$+TM@VC), and hence derive the overall change in resistance $\Delta R$ for a particular active site (TM@VC) as a function of the target molecule's concentration $C[X]$.

As a specific example of a chemical nanosensor we will consider the
functionalization of a (6,6) SWNT by nine different
 $3d$ TM atoms occupying three types of vacancies. For
six gas molecules (N$_{{2}}$, O$_{{2}}$, H$_{{2}}$O,
CO, NH$_{{3}}$, H$_{{2}}$S) we calculate the binding  
energy and change in conductance due to adsorption on each of the 27
TM sites. For a given type of TM functionalization, this allows us to
obtain the equilibrium coverage and change in conductance as a function
of the partial pressure of the ``target'' molecule in a background of
atmospheric air, as shown schematically in Figure \ref{Fig1}, and described in \cite{Juanma,Kirchberg}.

We will begin by providing a brief overview of the properties of our candidate sensing materials, specifically 3$d$ TM doped (6,6) SWNTs.  We will then discuss in detail in separate sections the four components of our method for modeling nanosensors: DFT calculations, kinetic modeling, NEGF results, and circuit theory, along with the results obtained for the TM functionalized SWNTs.  

\section{TM Doped SWNTs as Nanosensors}

To understand what a SWNT is \cite{Iijima,Rubio1,Dresselhaus,Harris,cnt_review}, one may imagine a sheet of graphene being rolled into a diminutive cylinder, which is very long ($\ell\sim 1$~cm) and narrow ($d\sim 1$~nm). The properties of the SWNT strongly depend on the way the graphene sheet is wrapped. To represent the way the graphene sheet is wrapped, two indices are used: ($n,m$), where $n$ and $m$ are integers. 

These two indices represent the number of unit cell vectors along the two principle directions in the crystal lattice of graphene, $\textbf{a}_1$ and $\textbf{a}_2$. 
For a ($n,m$) SWNT, the circumference $C$ is then given by 
\begin{equation}
 C = \|n \textbf{a}_1 + m \textbf{a}_2\|. 
\end{equation} 
The nanotube is metallic if $n-m = 3p$, where $p$ is an integer, and semiconducting otherwise.  If $n=m$ the tube is classified as an armchair SWNT, if $m=0$ the tube is called a zigzag SWNT, and otherwise the tube is known as chiral. Based on simple geometrical arguments, the diameter of a SWNT may be approximated using the relation
\begin{equation}
d
\approx \frac{a}{\pi}\sqrt{n^2+nm+m^2},
\end{equation}
where $a \approx 2.46$~\AA.  The strength, electronic structure, and high surface to volume ratio of SWNTs make them excellent materials for nanodevices.

SWNTs work remarkably well as detectors for small gas molecules, as demonstrated for both individual
SWNTs \cite{kong,collins,hierold,villalpando,rocha,Brahim} and SWNT networks \cite{morgan,cnt_networks,Goldoni}.  However, such sensitivity is most likely attributable to structural defects, vacancies, and junctions in the SWNT networks, as pristine SWNTs are inherently inert \cite{cnt_networks}.  Thus, to control a SWNT's performance as a chemical sensor, one must control the location, type, and number of these structural defects.  For this reason, the controlled doping and functionalization of SWNTs has become an area of increasing interest \cite{PaolaRevModPhys}, especially for substitutional doping with either N \cite{PaolaNdopedCNTs} or B \cite{PaolaBdopedCNTs}.

Previous studies have shown that SWNTs are highly sensitive to most molecules upon functionalization \cite{Fagan,Yagi,Yang,Chan,Yeung,Vo,Furst,Juanma,Krasheninnikov}.  However, the difficulty is determining which specific molecules are present.  In this study we show how changing the functionalization of the SWNT provides ``another handle'' for differentiating the SWNT's response to different gases/molecules.  In this way, one may not only detect that a molecule is present, but also differentiate which molecule is present.

Recent experimental advances now make the controlled doping of chirality selected SWNTs with metal atoms a possibility.  Specifically, these include (1) photoluminesence, Raman and XAS techniques for measuring the fraction of various SWNT chiralities in an enriched sample \cite{Kramberger07PRB,Ayala09PRB,DeBlauwe2010}; (2) the separation of SWNT samples by chirality using DNA wrapping \cite{Zheng03NM,Zheng03S,TU09N,Li07JOTACS}, chromatographic separation
\cite{Li07JOTACS,Fagan07JOTACS} and Density Gradient Ultracentrifugation
(DGU) \cite{Arnold06NN}; (3) SWNT resonators for measuring individual atoms of a metal vapor which adsorb on a SWNT \cite{Bachtold,ZettlMassSensor}; and (4) aberration corrected low energy ($<$50 keV) transmission electron microscopy (TEM) \cite{Chuanhong}.  The latter provides control over the formation of defects \emph{in situ} by adjusting the energy of the electron beam above and below the threshold energy for defect formation at a specific location on the SWNT.  These methods provide such a high level of control that it is now possible for experimentalists to take a specific SWNT chirality and dope the structure with individual metal atoms at a specified location.

At the same time, theorists are now able to embrace a ``bottom up'' approach to the design of nanosensors, harnessing the thermodynamics of self-assembly to find useful sensing devices \emph{in silico}.  With recent advances in both computational power and methodologies, theorists can now efficiently and accurately screen hundreds of candidate sensor designs using a combination of DFT for energetics of adsorption and stability, and NEGF methodologies for the electrical response \cite{Juanma}.

\begin{figure}[!tb]
\centering
\includegraphics[width=0.7\textwidth]{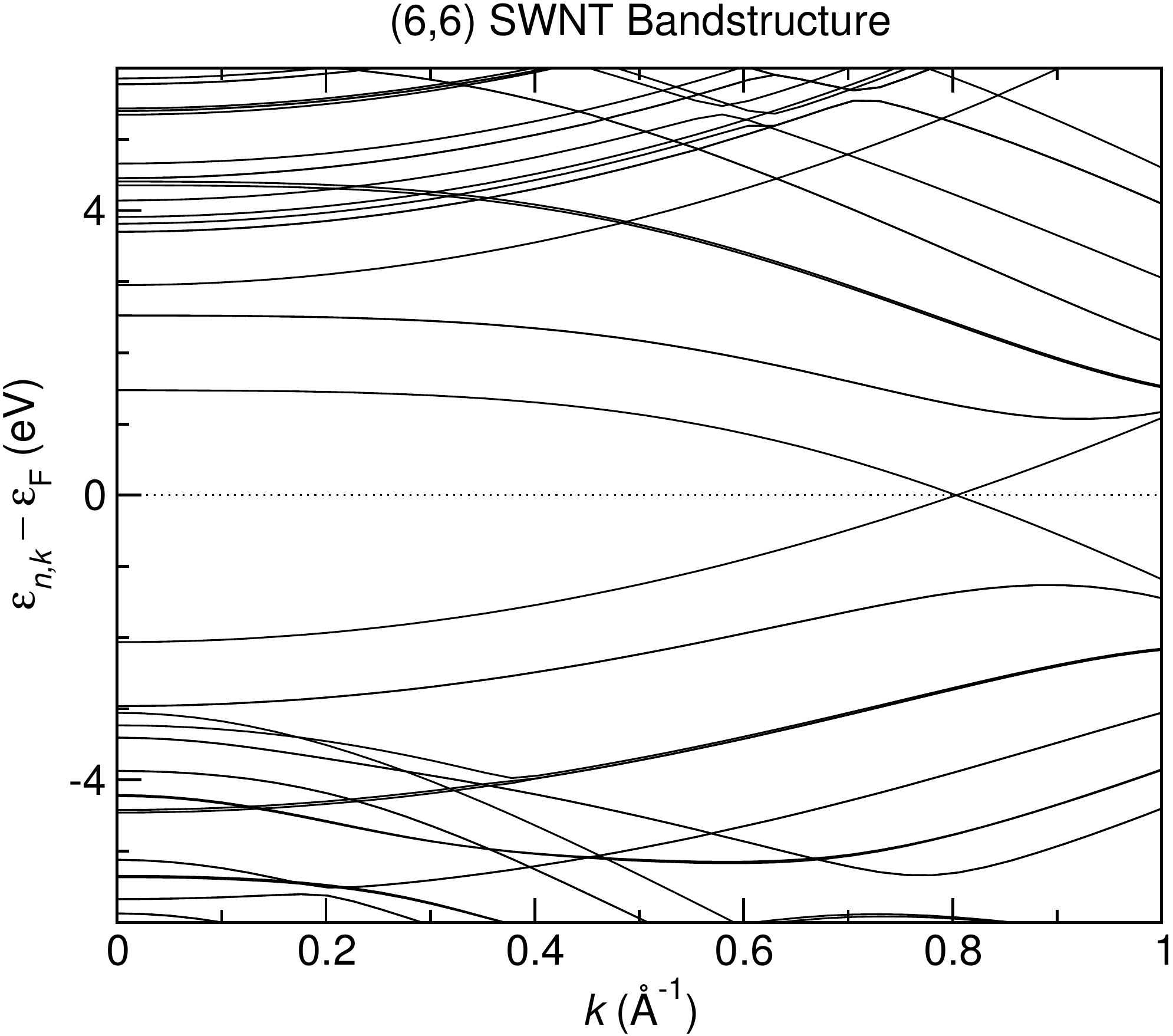}
\caption{Band energies $\varepsilon_{n,k}$ relative to the Fermi level $\varepsilon_{\mathrm{F}}$ in eV versus momentum $k$ in \AA$^{-1}$ for a pristine (6,6) metallic armchair SWNT.}\label{bandstructure}
\end{figure}

In this chapter, we have used nine of the ten $3d$ TMs as candidates for substitutional doping of a (6,6) metallic armchair SWNT. The (6,6) SWNT was chosen as it is small enough to be efficiently calculated (24 atoms in the minimal unit cell, 144 TM@MV or 143 atoms for the TM@DVI, TM@DVII), and at the same time experimentally realizable \cite{Dresselhaus}.
We repeat the minimal unit cell 6 times to ensure convergence of the
Hamiltonian to its bulk values at the borders of the cell.  In other
words, the unit cell is large enough that structural and electronic
changes due to the TM in the vacancy are not felt at the
unit cell boundary.  The radius of the (6,6) SWNT is about 4.1 \AA, while in experimental samples of chirality sorted tubes, the radius is typically around 7 \AA.  

\begin{figure}[!t]
\centering
\includegraphics[width=\textwidth]{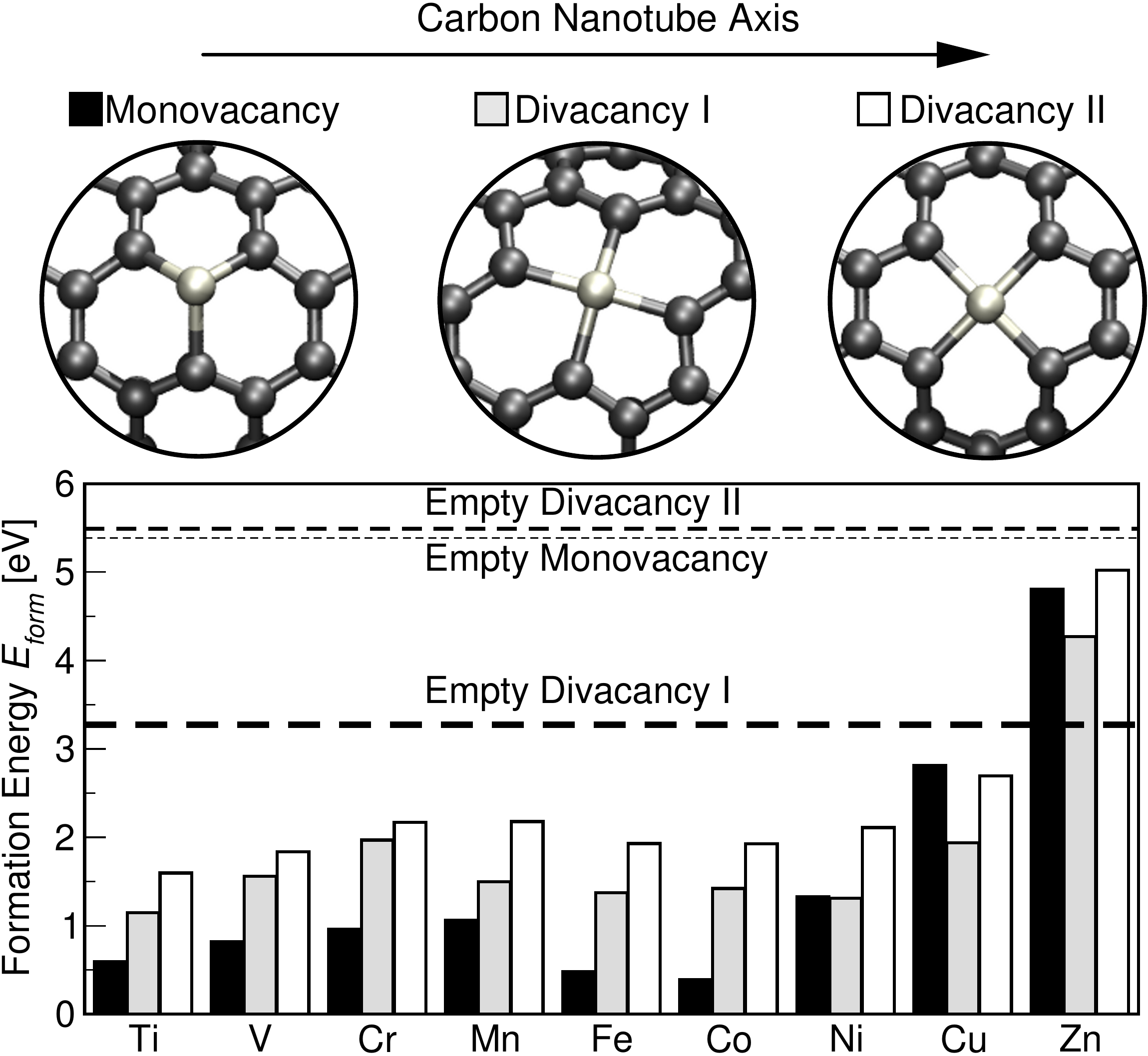}
\caption{Structural schematics and formation energies $E_{\mathrm{form}}$ in eV relative to a pristine (6,6) SWNT and a physisorbed transition metal atom for a $3d$
transition metal occupied monovacancy (TM@MV, black), divacancy I (TM@DVI, grey), and divacancy II (TM@DVII, white) in a (6,6) SWNT. Vacancy formation energies
for an empty monovacancy (- - - - -), divacancy I ({\Large{\bf{-- -- -- --}}}), and divacancy II (\textbf{-- -- -- --}) are shown for comparison ({Garc{\'{\i}}a-Lastra} et~al.~2010)
.}\label{Stability}
\end{figure}

However, in the $-1$ to 1 eV energy range of interest, for metallic
tubes, the DOS is flat (2 eV$^{-1}$), and has two open eigenchannels, so
that T$(\varepsilon) = 2$.  This is clearly seen from the electronic band structure of the (6,6) SWNT, which is shown in Figure \ref{bandstructure}.  In this way, the (6,6) SWNT results also provide a qualitative understanding of the response of larger metallic tubes.  Further, for metallic armchair ($n,n$) SWNTs both the transmission T$(\varepsilon)$ and DOS are completely flat near the Fermi level.
This makes it much easier to differentiate the affect of the
TM dopant, and the adsorbed molecule on the transmission
and DOS.

The $3d$ TM atoms have been chosen as candidate dopants for several reasons.  First, the $3d$ TM atoms are smaller for a given electronic
configuration than the $4d$ and $5d$ TMs, and are expected
to fit best in the structure of a SWNT, with the least strain.  Even
so, the TM atoms are typically pushed somewhat above the
SWNT surface, due to both the strain on the graphene lattice, and the
difference in atomic radii.  These radii vary from the largest at 1.76
\AA\ for Ti, to the smallest at 1.42 \AA\ for Ni \cite{Ashcroft}.  These
are all between 2 and 3 times the atomic radius of C of 0.67 \AA.
Also, the electronic properties of a TM dopant are
mostly determined by the filling of the $d$-band, and should be quite
similar for the $4d$ and $5d$ counterparts of the $3d$ TMs
studied herein.

The monovacancy, divacancy I and divacancy II were considered here since 
these are the three most stable vacancies, as depicted schematically in Figure \ref{Stability}.  Comparing divacancy I and
divacancy II results shows the effect of strain on the electronic
properties of the active site, since for the divacancy II the strain
is evenly distributed over the four bonds, while for the divacancy I
the strain is mostly on two of the four bonds, as seen in Figure \ref{Stability}.  For the monovacancy,
the TM dopant is three fold coordinated, while for the
divacancies, the TM dopant is four-fold coordinated.
This explains why in general we find molecules are more strongly
bonded to TMs in a monovacancy compared to either of the
divacancies.  

For a strongly binding target molecule, such as CO, a
divacancy provides the most suitable active site, as the site will not
be completely saturated/covered, and a significant fraction of
empty sites should always be present.  On the other hand, for a weakly binding molecule, such as NH$_3$, a monovacancy is more suitable, as there will then be a measurable change in coverage with molecule concentration.  

%
%

\section{Density Functional Theory}

\begin{figure}[!h]
\centering
\includegraphics[scale=0.5]{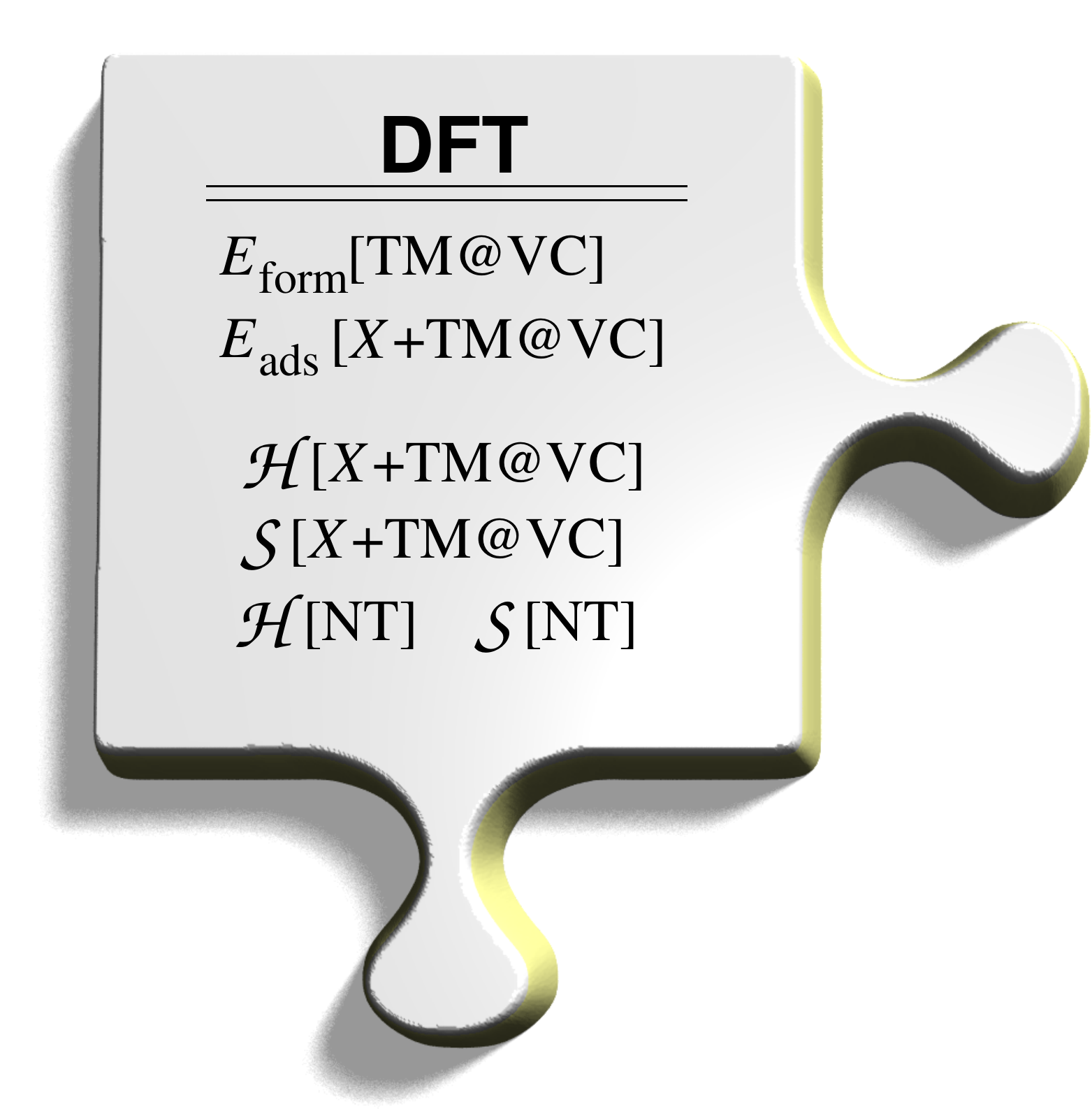}
\caption{From density functional theory (DFT) we obtain the formation energies $E_{\mathrm{form}}$ of an active site (a 3$d$ transition metal occupied vacancy (TM@VC)), the adsorption energies $E_{\mathrm{ads}}$ for species $X$ on the active site ($X$+TM@VC), and the Hamiltonian $\mathpzc{H}$ and overlap matrices $\mathpzc{S}$ for the scattering region with $X$ adsorbed ($X$+TM@VC) and the leads (a pristine (6,6) SWNT (NT)).}\label{DFTPiece}
\end{figure}

At its root, DFT is an exact reformulation of quantum mechanics in terms of the electron density \cite{Kohn-Sham,Parr,DFTMarques,DFT2011}.  In this way it is a quantum mechanical  method for calculating the ground state energy for many-body systems. From this theory, we may calculate formation energies $E_{\mathrm{ads}}$, adsorption energies $E_{\mathrm{ads}}$, and a system's Hamiltonian $\mathcal{H}$ and overlap $\mathcal{S}$ matrices, which form the first ``puzzle piece'' of our chemical nanosensor model, as shown in Figure~\ref{DFTPiece}. 

The reason why DFT is used is that it simplifies a problem in terms of the many electron wave function $\Psi(\textbf{r}_1,\cdots,\textbf{r}_N)$ into another problem, which is only in terms of the
electron density $\rho(\textbf{r}_1)$. This is an important advantage because it reduces the coordinates from 3$N$ to just 3 $(x,y,z)$.
\begin{eqnarray*}
\Psi(\textbf{r}_1,\ldots,\textbf{r}_N) &\rightarrow& 3N\ \mathrm{coordinates},\\
\varepsilon_0[\rho] \rightarrow \rho(\textbf{r}_1) &\rightarrow& 3\ \mathrm{coordinates}.
\end{eqnarray*}
This is based on the Hohenberg Kohn theorem which states that for a given external potential
the ground state energy is a unique functional of the electron density.
\begin{equation}
\varepsilon_0[\rho] = F[\rho] + \int d\textbf{r} \rho(\textbf{r}) v_{ext}(\textbf{r}), 
\end{equation}
where
\begin{equation}
F[\rho] = T[\rho] + V_{ee}[\rho].
\end{equation}
Here $T[\rho]$ is the kinetic energy, $V_{ee}[\rho]$ describes the electron-electron interaction and $v_{ext}$ is
the external potential. However, an exact form for $F[\rho]$ in terms of the density is unknown, so
we approximate it by
\begin{equation}
F[\rho] = T[\rho] + \frac{1}{2}\int\!\!\!\int d\textbf{r}d\textbf{r}' \frac{\rho(\textbf{r})\rho(\textbf{r}')}{\|\textbf{r}-\textbf{r}'\|} - \int d\textbf{r} \rho(\textbf{r})v_{xc}[\rho](\textbf{r}).
\end{equation}
Here $v_{xc}$ is the exchange correlation potential that gives the correct electron-electron interaction $V_{ee}$. $v_{xc}$ has corrections to the Coulomb potential in order to fulfill the Pauli exclusion principle, self interactions corrections, and electron-electron correlations. 

The most common implementation of DFT uses the Kohn-Sham method. The procedure for this method is to assume a form for $v_{xc}$, and then solve the non-interacting Schr\"{o}dinger equation
\begin{equation}
-\frac{\hslash^2}{2m}\nabla^2\psi_i + v(\textbf{r})\psi_i = \varepsilon_i\psi_i, \label{Psi}
\end{equation}
with the effective potential $v$ given by
\begin{equation}
v[\rho](\textbf{r}) = \int d\textbf{r}' \frac{\rho(\textbf{r}')}{\|\textbf{r} - \textbf{r}'\|} + v_{xc}[\rho] + v_{ext}(\textbf{r}).\label{v}
\end{equation}
This potential depends on the density 
\begin{equation}
\rho = \sum_{i=0}^{N/2} \psi_i^*\psi_i,\label{rho}
\end{equation}
which in turn depends on the non-interacting Kohn-Sham wave functions $\psi_i$, where $N$ is the number of electrons.  We may thus obtain the wave functions and density by solving Eqns.~(\ref{Psi}), (\ref{v}), and (\ref{rho}) in a self consistent manner.  This Kohn-Sham self-consistency procedure is at the heart of DFT \cite{Kohn-Sham}.  

In this work, all total energy calculations and structural optimizations have been
performed within the real-space density functional theory (DFT) code \textsc{gpaw}  \cite{GPAW,GPAWRev} which
is based on the projector augmented wave method (PAW) \cite{Blochl}.  This allows one to use a coarse grid, with ``smoothed'' pseudo-wave functions within the core, which describes the valence and conduction states quite well, to calculate the pseudo-density and pseudo-wave functions self-consistently.  At the same time, it allows access to the full all-electron densities and wave functions, by projecting back onto the core wave functions on a finer grid.  We find a grid spacing
of 0.2~\AA~provides a quite accurate representation of both the density and wave functions. For the purposes of energetics of molecules and solids, the Perdew-Burke-Ernzerhof implementation of the GGA for the 
exchange correlation functional \cite{PBE} works quite well. Spin
polarization has been taken into account in all calculations.  This is necessary, since $3d$ TM atoms often have a magnetic moment, as do molecules in the atmosphere such as O$_2$.

Pristine SWNTs are known to
be chemically inert -- a property closely related to their high stability. As a consequence, only radicals bind strong enough to the SWNT to notably affect its electrical
properties \cite{RevModPhys.79.677,hierold,Valentini2004356,zanolli:155447,Juanma}. To make SWNTs attractive for sensor
applications thus requires some kind of functionalization, e.g. through
doping or decoration of the SWNT sidewall \cite{Fagan,Yagi,Yang,PaolaNdopedCNTs,Chan,Yeung,Vo,Furst,JuanmaPRL,PaolaBdopedCNTs,Krasheninnikov,PaolaRevModPhys}. Ideally, this type of functionalization
could be used to control not only the reactivity of the SWNT but also the
selectivity towards specific chemical species.

Metallic doping of a (6,6) SWNT has been modeled in a supercell containing six repeated
minimal unit cells along the SWNT axis (dimensions: 15 \AA$\times$15 \AA$\times$14.622 \AA).
For this size of supercell a $\Gamma$-point sampling of the Brillouin zone was found to be
sufficient to obtain reasonably converged adsorption and formation energies.  It should also be noted that for transport calculations within the NEGF method, it is implicitly assumed that the interaction between atoms near opposite ends of the unit cell are negligible, and any $k$-point dependence along the transport direction is neglected.  

We define the formation energy for creating a
vacancy (VC) occupied by a TM atom using
the relation
\begin{equation}
E_{\mathrm{form}}[\mathrm{TM}@\mathrm{VC}] = E[\mathrm{TM}@\mathrm{VC}] +
n E[\mathrm{C}] - E[\mathrm{TM+NT}],
\end{equation}
where $E$[TM@VC] is the total energy of a TM atom
occupying a vacancy in the nanotube, $n$ is the number of carbon atoms
removed to form the vacancy, $E$[C] is the energy per carbon atom in
a pristine nanotube, and $E$[TM+NT] is the total energy of the pristine nanotube with a 
physisorbed TM atom. The vacancies considered herein are the monovacancy and two divacancies shown in Figure \ref{Stability}.  The energy required to form an empty vacancy is obtained from 
\begin{equation}
E_{\mathrm{form}}[\mathrm{C}@\mathrm{VC}] = E[\mathrm{VC}] + n
E[\mathrm{C}] - E[\mathrm{NT}], 
\end{equation}
where $E$[VC] is the total energy of the nanotube with a vacancy of $n$ atoms. 

The calculated formation energies for the $3d$ TMs are
shown in Figure \ref{Stability}. From the horizontal lines we see that both
divacancies are more stable than the monovacancy. This may be
attributed to the presence of a two-fold coordinated C atom in the
monovacancy, while all C atoms remain three-fold coordinated in the
divacancies.  When a TM atom occupies a vacancy, the
strongest bonding to the C atoms is through its $d$
orbitals~\cite{Griffith}. For this reason, Cu and Zn, which both have
filled $d$-bands, are rather unstable in the SWNT. For the remaining
metals, adsorption in the monovacancies leads to quite stable
structures. This is because the three-fold coordination of the C atoms
and the SWNT's hexagonal structure are recovered when the metal atom is
inserted. On the other hand, metal adsorption in divacancies is
slightly less stable because of the resulting pentagon defects, depicted in the
upper panel in Figure \ref{Stability}.  A similar behavior has been reported
by Krasheninnikov \emph{et al.} for TM atoms in graphene
\cite{Krasheninnikov}. 

\begin{table}
\caption{Adsorption energies $E_{\mathrm{ads}}$ in eV for N$_2$, O$_2$, H$_2$O, CO, NH$_3$, and H$_2$S on a $3d$ transition metal occupied monovacancy (TM@MV).}\label{TM@MV}
\begin{tabular}{l|ccccccccc}
& Ti & V & Cr & Mn & Fe & Co & Ni & Cu & Zn \\\hline
N$_2$ & -0.61 & -0.76 & -0.57 & -0.70 & -0.77 & -0.60 & -0.49 & -0.34 & -0.51\\
O$_2$ & -3.16 & -3.39 & -2.61 & -2.57 & -2.17 & -1.88 & -2.00 & -1.02 & -0.90\\
H$_2$O & -1.09 & -1.08 & -0.97 & -0.89 & -0.84 & -0.64 & -0.54 & -0.52 & -0.62\\
CO & -0.89 & -1.21 & -1.06 & -1.38 & -1.54 & -1.23 & -1.16 & -0.97 & -1.07\\
NH$_3$ & -1.39 & -1.46 & -1.35 & -1.32 & -1.31 & -1.03 & -0.94 & -0.89 & -1.12\\
H$_2$S & -0.78 & -0.88 & -0.77 & -0.78 & -0.88 & -0.58 & -0.69 & -0.63 & -0.67
\end{tabular}
\end{table}

\begin{table}
\caption{Adsorption energies $E_{\mathrm{ads}}$ in eV for N$_2$, O$_2$, H$_2$O, CO, NH$_3$, and H$_2$S on a $3d$ transition metal occupied divacancy I (TM@DVI).}\label{TM@DVI}
\begin{tabular}{l|ccccccccc}
& Ti & V & Cr & Mn & Fe & Co & Ni & Cu & Zn \\\hline
N$_2$ & -0.58 & -0.65 & -0.65 & -0.38 & -0.46 & -0.63 & -0.15 & -0.01 & -0.01\\
O$_2$ & -2.72 & -3.41 & -2.74 & -1.92 & -1.67 & -1.42 & -0.61 & -0.06 & -0.12\\
H$_2$O & -1.21 & -1.19 & -0.79 & -0.49 & -0.51 & -0.33 & -0.19 & -0.05 & -0.25\\
CO & -0.80 & -1.11 & -1.14 & -1.06 & -1.62 & -1.90 & -1.10 & -0.12 & -0.04\\
NH$_3$ & -1.51 & -1.58 & -1.11 & -0.89 & -0.91 & -0.75 & -0.40 & -0.20 & -0.53\\
H$_2$S & -0.87 & -0.78 & -2.13 & -0.50 & -0.64 & -0.59 & -0.31 & -0.04 & -0.17
\end{tabular}
\end{table}

\begin{table}
\caption{Adsorption energies $E_{\mathrm{ads}}$ in eV for N$_2$, O$_2$, H$_2$O, CO, NH$_3$, and H$_2$S on a $3d$ transition metal occupied divacancy II (TM@DVII).}\label{TM@DVII}
\begin{tabular}{l|ccccccccc}
& Ti & V & Cr & Mn & Fe & Co & Ni & Cu & Zn \\\hline
N$_2$ & -0.50 & -0.62 & -0.41 & -0.38 & -0.60 & -0.39 & -0.07 & -0.02 & -0.11\\
O$_2$ & -2.27 & -2.64 & -2.35 & -2.17 & -1.48 & -1.31 & -0.53 & -0.09 & -0.18\\
H$_2$O & -1.15 & -1.20 & -0.74 & -0.64 & -0.49 & -0.55 & -0.16 & -0.05 & -0.17\\
CO & -0.78 & -1.10 & -1.05 & -0.96 & -1.49 & -1.30 & -0.82 & 0.02 & 0.01\\
NH$_3$ & -1.45 & -1.47 & -1.14 & -1.09 & -0.99 & -0.68 & -0.23 & -0.13 & -0.42\\
H$_2$S & -0.80 & -0.83 & -0.62 & -0.62 & -0.56 & -0.50 & -0.10 & -0.01 & -0.10
\end{tabular}
\end{table}

The adsorption energies for N$_2$, O$_{\mathrm{2}}$,
H$_{\mathrm{2}}$O, CO, NH$_{\mathrm{3}}$, and H$_{\mathrm{2}}$S on the metallic site of the doped (6,6) SWNTs are shown in
Tables~\ref{TM@MV}, \ref{TM@DVI}, and \ref{TM@DVII} for the TM occupied monovacancy, divacancy I, and divacancy II, respectively.  The adsorption energy of a molecule $X$ is defined by 
\begin{equation}
E_{\mathrm{ads}}[X\mathrm{+TM@VC}] = E[X\mathrm{+TM@VC}] - E[X] - E[\mathrm{TM@VC}], 
\end{equation}
where $E$[$X$@TM@VC] is the total energy of molecule $X$ on a
TM atom occupying a vacancy, and $E[X]$ is the gas phase
energy of the molecule. 

From the adsorption energies shown in Tables \ref{TM@MV}, \ref{TM@DVI}, and \ref{TM@DVII}, we see that
the earlier TMs tend to bind the adsorbates stronger
than the late TMs. The latest metals in the series (Cu and Zn) bind adsorbates rather weakly in the divacancy structures. We
also note that O$_2$ binds significantly stronger than any of the
three target molecules on Ti, V, Cr, and Mn (except for Cr in the
divacancy I where H$_2$S is found to spontaneously dissociate).  Active sites
containing these metals are therefore expected to be completely
passivated if oxygen is present in the background. Further, we find H$_2$O is rather weakly bound to most of the active sites.  This ensures that these types of sensors are robust against changes in humidity.

\section{Kinetic Modeling}

\begin{figure}[!h]
\centering
\includegraphics[scale=0.5]{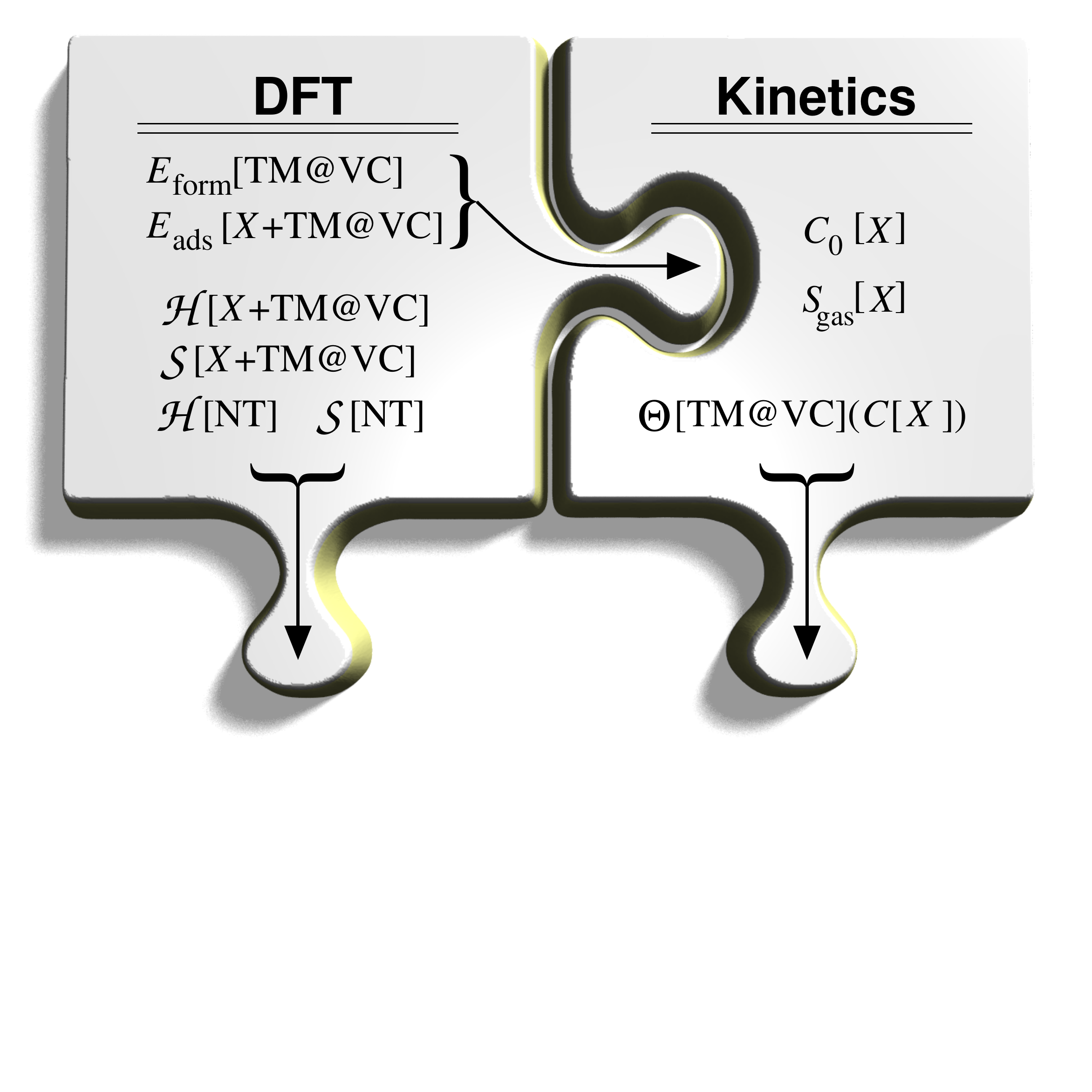}
\caption{The energetics from density functional theory (DFT) are inputted to a kinetic model for the active site in thermodynamic equilibrium with concentrations $C_0$, and gas phase entropies $S_{\mathrm{gas}}$,  for each species $X$ taken from experiment (Lide~2006--2007)
, to obtain the fractional coverage $\Theta$ per active site (a 3$d$ transition metal occupied vacancy TM@VC) as a function of the target molecule's concentration $C[X]$.}\label{Kinetics}
\end{figure}

The next ``piece of the puzzle'' for our chemical nanosensor model is the description of the coverage of our active sites by both target and background molecules using kinetic theory, as depicted in Figure \ref{Kinetics}.  

\begin{table}
\caption{Equilibrium atmospheric concentrations $C_0[X]$ at 4\% humidity and gas phase entropies $S_{\mathrm{gas}}[X]$ in meV/K at $T=300$ K where $X$ is one of the background molecules N$_2$, O$_2$, and H$_2$O or one of the target molecules CO, NH$_3$, and H$_2$S. Experimental gas phase entropies are taken from Ref.~(Lide~2006--2007)
.}\label{Table1}
\begin{tabular}{lrclrc}
$X$  & $C_0[X]$  & $S_{\mathrm{gas}}[X]$&$X$  & $C_0[X]$  & $S_{\mathrm{gas}}[X]$\\\hline
N$_{{2}}$  & 74.96\% & 1.988 meV/K&CO            & 96.00 ppb &2.050 meV/K\\
O$_{{2}}$  & 20.11\% & 2.128 meV/K&NH$_{{3}}$ & 16.32 ppb &2.000 meV/K\\
H$_{{2}}$O & 4.00\% & 1.959 meV/K&H$_{{2}}$S & 0.96  ppb &2.136 meV/K\\
\end{tabular}
\end{table}

For a sensor containing hundreds, if not thousands of sites, the fractional coverage of sites is reasonably described by the fractional coverage in thermodynamic equilibrium $\Theta$ at standard temperature and pressure in terms of the target molecule concentration.   The adsorption reaction for a species $X$ on an empty active site * is then
\begin{equation}
X + * \leftrightarrow X^*.
\end{equation}
Under equilibrium conditions, the rate of adsorption and desorption of species $X$ must be equal, so that
\begin{equation}
r_{\mathrm{ads}}[X] = k_+[X] \Theta[*] C[X] - k_-[X] \Theta[X] = 0,\label{rate}
\end{equation}
where $r_{\mathrm{ads}}$ is the rate of the adsorption reaction, $k_+$ is the forward rate constant, $k_-$ is the backward rate constant, $C$ is the atmospheric concentration, $\Theta[*]$ is the fraction of active sites which are empty, and $\Theta[X]$ is the fraction of active sites which are covered by species $X$.  

Using Eqn.~(\ref{rate}) we may express the coverage of active sites by species $X$ in terms of the ratio of forward and backward rate constants $K = k_+/k_-$ and $\Theta[*]$ as
\begin{equation}
\Theta[X] = K[X]\Theta[*]C[X].\label{ThetaX}
\end{equation}
From kinetic theory, the ratio of rate constants may be obtained from the DFT adsorption energy $E_{\mathrm{ads}}$ as
\begin{equation}
K[X] = \exp \left[\frac{ -E_{\mathrm{ads}}[X] - T S_{{gas}}[X]}{k_B T}\right],
\end{equation}
where $k_B$ is Boltzmann's constant, $T$ is the temperature, and $S_{{gas}}[X]$ is the gas phase entropy of species $X$, as given in Table~\ref{Table1} \cite{CRCHandbook}.

Since Eqn.~(\ref{rate}) is valid for all species in the set of background molecules $\mathcal{B}$, so long as our description of the background is sufficient, if we sum up all the fractional coverages for each species, and the fraction of sites which are empty, we should obtain unity.  More specifically,
\begin{equation}
\Theta[*] + \Theta[X] + \sum_{Y\in\mathcal{B}}\Theta[Y] = 1.\label{unity}
\end{equation}

By simple repeated substitution of Eqn.~(\ref{ThetaX}) into Eqn.~(\ref{unity}), we may express the fractional coverage of active sites by species $X$ in terms of the ratios of forward and backward rate constants, and concentrations in atmosphere as
 \begin{equation}
\Theta[X] = \frac{K[X] C[X]}{1 + K[X]C[X] + \sum_{Y \in \mathcal{B}} K[Y] C[Y]}.
\end{equation}

\begin{table}
\caption{Logarithm of the change in fractional coverage with atmospheric concentration of the target molecule $\log\left[\frac{\partial \Theta[X]}{\partial C[X]}\right]$, for CO, NH$_3$, and H$_2$S on a $3d$ transition metal occupied monovacancy (TM@MV).}\label{ActTM@MV}
\begin{tabular}{l|ccccccccc}
& Ti & V & Cr & Mn & Fe & Co & Ni & Cu & Zn \\\hline
CO & -86 & -82 & -58 & -44 & -22 & -23 & -30 & -7 & +9\\
NH$_3$ & -66 & -72 & -46 & -46 & -31 & -30 & -38 & -2 & +12\\
H$_2$S & -91 & -96 & -70 & 68 & -49 & -49 & -50 & -14 & -7
\end{tabular}
\end{table}

\begin{table}
\caption{Logarithm of the change in fractional coverage with atmospheric concentration of the target molecule $\log\left[\frac{\partial \Theta[X]}{\partial C[X]}\right]$, for CO, NH$_3$, and H$_2$S on a $3d$ transition metal occupied divacancy I (TM@DVI).}\label{ActTM@DVI}
\begin{tabular}{l|ccccccccc}
& Ti & V & Cr & Mn & Fe & Co & Ni & Cu & Zn \\\hline
CO & -72 & -87 & -60 & -31 & +1 & +11 & +13 & -19 & -22\\
NH$_3$ & -44 & -68 & -60 & -37 & -27 & -28 & -10 & -15 & -2\\
H$_2$S & -71 & -101 & -22 & -54 & -38 & -36 & -15 & -23 & -18
\end{tabular}
\end{table}

\begin{table}
\caption{Logarithm of the change in fractional coverage with atmospheric concentration of the target molecule $\log\left[\frac{\partial \Theta[X]}{\partial C[X]}\right]$, for CO, NH$_3$, and H$_2$S on a $3d$ transition metal occupied divacancy II (TM@DVII).}\label{ActTM@DVII}
\begin{tabular}{l|ccccccccc}
& Ti & V & Cr & Mn & Fe & Co & Ni & Cu & Zn \\\hline
CO & -56 & -58 & -48 & -45 & +3 & +2 & +8 & -23 & -24\\
NH$_3$ & -29 & -43 & -44 & -39 & -16 & -21 & -14 & -18 & -7\\
H$_2$S & -56 & -69 & -66 & -59 & -34 & -30 & -21 & -22 & -21
\end{tabular}
\end{table}

To measure the ``sensitivity'' of an active site to the concentration of a target molecule, one should consider how the coverage of the active site by  the target molecule changes with the concentration of the species in the atmosphere.  It is only when there are changes in the active site's coverage, that through the resulting changes in the sensing property, the nanosensor may be useful for a given species' detection.  Specifically, we find the logarithm of the change in coverage with concentration of the target molecule, i.e. the derivative
\begin{equation}
\log\left(\frac{\partial \Theta[X]}{\partial C[X]}\right) = \log \left(\frac{\Theta[X]}{C[X]}\left(1-\Theta[X]\right)\right),
\end{equation}
acts as a reasonable ``descriptor'' for an active site's sensitivity to species $X$.

\begin{figure}[!th]
\centering
\includegraphics[width=\textwidth]{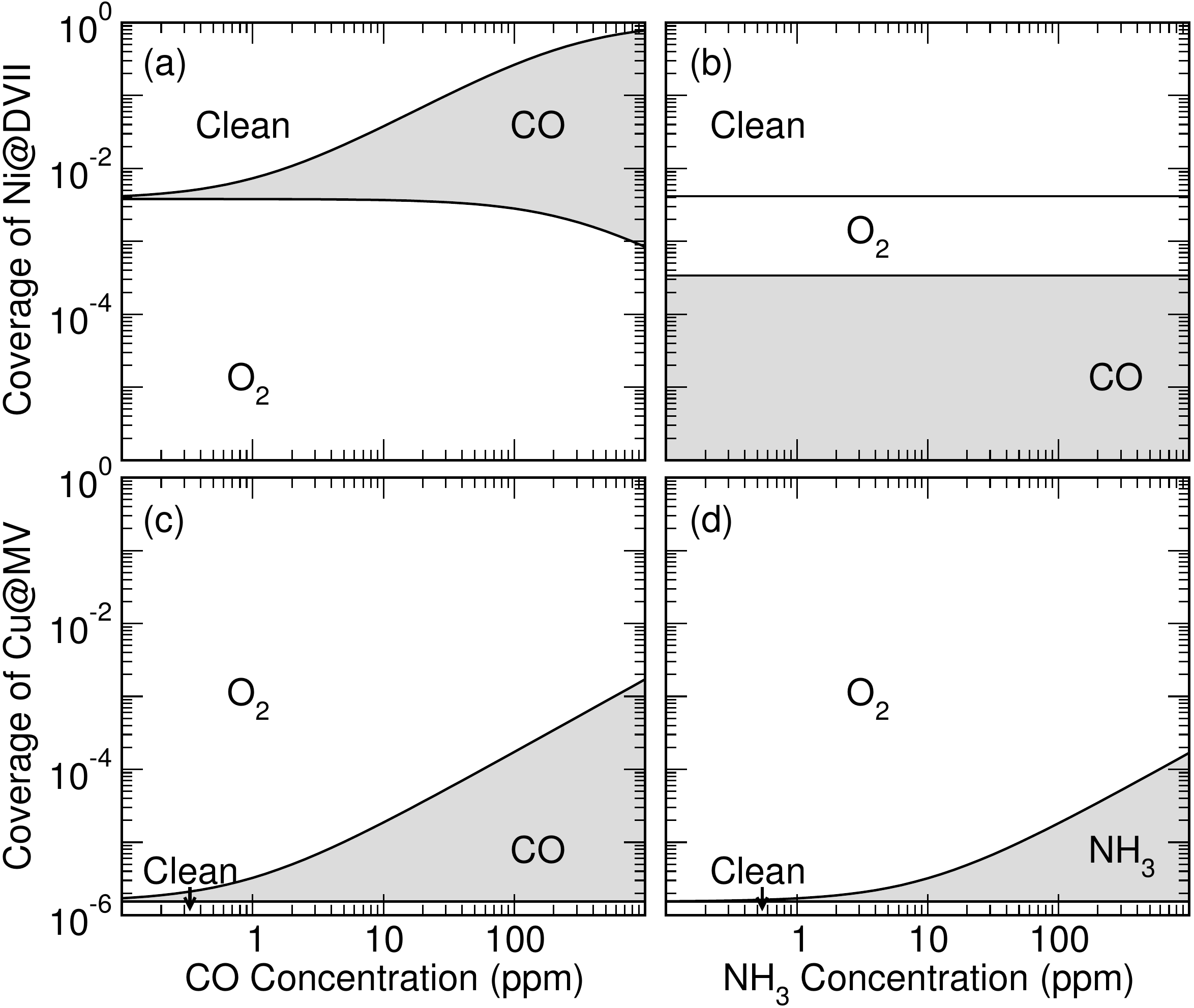}
\caption{%
 Fractional coverage $\Theta$ in thermal equilibrium of (a,b) Ni in a divacancy II (Ni@DVII) and (c,d) Cu in a monovacancy (Cu@MV) versus (a,c) CO concentration and (b,d) NH$_{{3}}$ concentration in a background of air at room temperature and 1 bar of pressure (Mowbray et~al.~2010)
.}
\label{Fig3}
\end{figure}

As shown in Table \ref{ActTM@MV}, in the monovacancy, only the relatively inert Cu and Zn active sites are at all sensitive to the target molecules, specifically for CO and NH$_3$.  This is due to the binding energy of O$_2$, which is $\gtrsim 2$~eV for all other TM atoms occupying monovacancies, but only $\sim 1$~eV for both Cu and Zn, as shown in Table \ref{TM@MV}.  This leaves ``clean'' Cu@MV and Zn@MV active sites, where CO or NH$_3$ may adsorb.

On the other hand, for both the divacancies shown in Tables \ref{ActTM@DVI} and \ref{ActTM@DVII} we find Fe, Co, and Ni are sensitive to CO.  We attribute this to the similarity in binding energies for O$_2$ and CO on these sites, which are between 0.5 and 1.6 eV, as shown in Tables \ref{TM@DVI} and \ref{TM@DVII}.  We again find Ti, V, Cr, and Mn are completely oxidized, as was the case in the monovacancy, making these sites inactive, while for Cu and Zn the binding energies of the target molecules are too weak for them to adsorb on these sites.

In Figure \ref{Fig3} we see the coverages of both Ni and Cu active sites as a function of CO and NH$_{{3}}$ concentration.  Clearly, both Cu and Ni active sites will be sensitive to the adsorption of CO.  On the other hand, the coverage of the Cu active sites is sensitive to NH$_{{3}}$ concentration, while Ni active sites are not.  This suggests that by combining the response of Cu and Ni active sites, we may obtain a multifunctional sensor for both CO and NH$_{{3}}$.

\section{Non-Equilibrium Green's Function Methodology}

\begin{figure}[!h]
\centering
\includegraphics[scale=0.5]{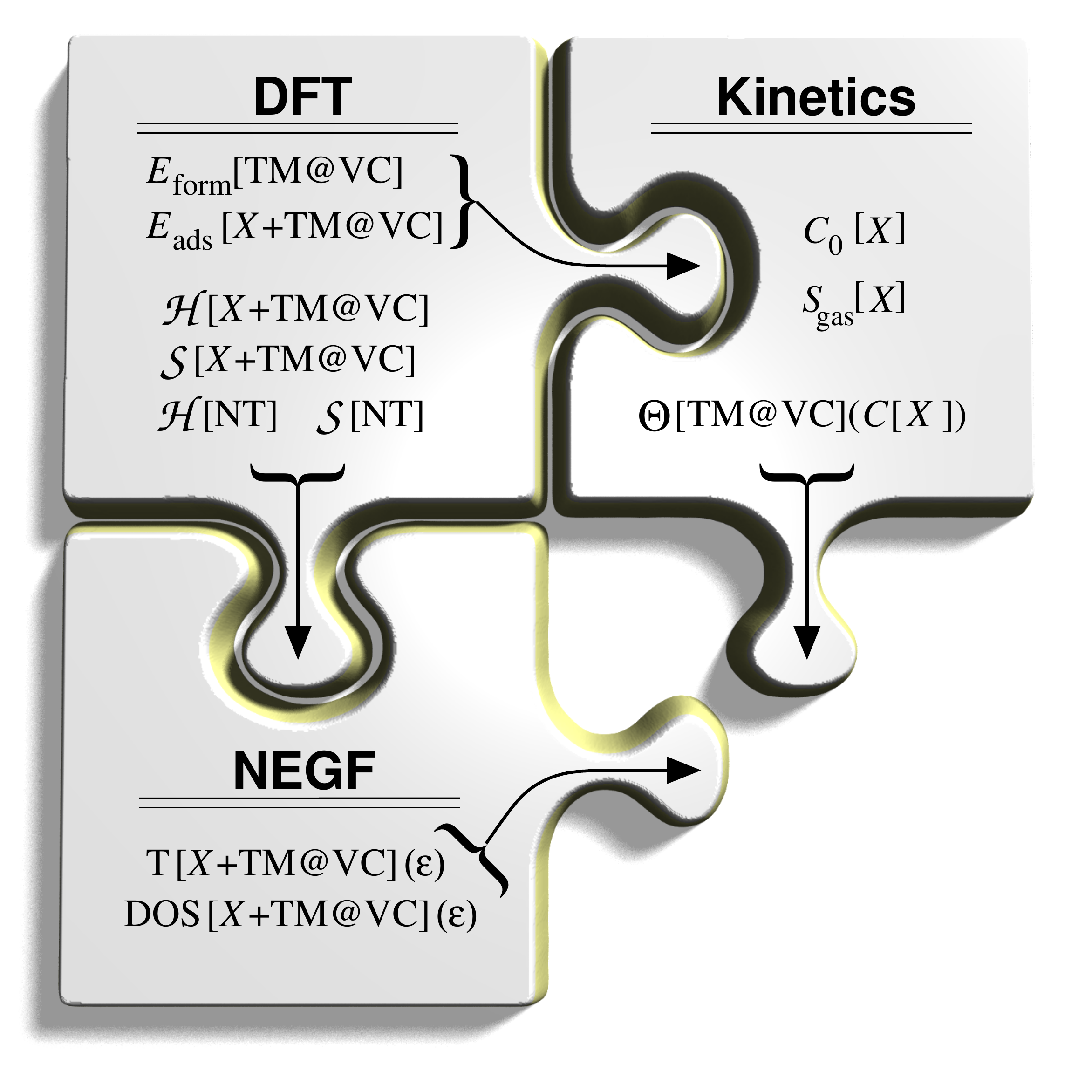}
\caption{The Hamiltonian $\mathpzc{H}$ and overlap $\mathpzc{S}$ matrices obtained from density functional theory (DFT) are used within the non-equilibrium Green's function (NEGF) methodology to obtain the density of states (DOS) and transmission probability T for each species $X$ on an active site ($X$+TM@VC), as a function of energy $\varepsilon$.
}\label{NEGF}
\end{figure}

The third ``puzzle piece'' needed to model our chemical nanosensor is the description of the sensing property, specifically the change in resistance of the sensor using the NEGF method, as depicted in Figure \ref{NEGF}.
Although molecules may adsorb on the active sites, the sensing property of the SWNT resistance must change significantly for the sensor to be useful.  In the following sections we show both the density of states (DOS) and transmission probability T($\varepsilon$)  for an electron of a given energy $\varepsilon$ to travel past an active site as calculated within the NEGF method.

Since the energy levels of a given molecule may be thought of as its electronic ``finger print'',  so long as there is sufficient binding between the molecule and active site, this ``finger print'' is evidenced in the DOS of the system.  Furthermore, the peaks of the DOS due to the molecule tend to form Fano anti-resonances \cite{Furst} in the transmission, as electrons are scattered off these states.  In this way, any adsorbed molecule tends to leave its ``finger print'' in the transmission probability through an active site.  For this reason, conductance/resistance is generally an effective sensing property.

The NEGF methodology \cite{Datta} is based on the Green's function, which is dependent on two space-time coordinates. The NEGF method can be applied to both extended and infinite systems and it can handle strong external fields without being perturbed. The approximations made in the NEGF method can be chosen in order to satisfy macroscopic conservation laws. An interesting feature of the NEGF method is that memory effects in transport that occur due to electron-electron interactions can be analyzed.


As depicted schematically in Figure \ref{NEGF}, the NEGF method requires the Hamiltonian $\mathcal{H}$ and overlap $\mathcal{S}$ matrices for both the clean and doped SWNT structures.  The NEGF calculation is then performed using as ``leads'' the clean SWNT Hamiltonian, with a ``scattering region'' defined by the doped SWNT Hamiltonian.  Within this representation, the states which form the Hamiltonian must be in some way localized, so that the Hamiltonian of the scattering region is ``converged'' to the Hamiltonian of the leads near the unit cell boundaries.  This may be accomplished through the use of Wannier functions to localize periodic wave functions, or a separate calculation with a locally centered atomic orbital (LCAO) basis set \cite{BenchmarkPaper}.  

In this work, transport calculations for the optimized structures have been performed  with an
electronic Hamiltonian obtained from the \textsc{siesta} code
 \cite{SIESTA} in a double zeta polarized (DZP) LCAO basis set.  Previous studies have shown that a DZP basis set is sufficient to converge the Hamiltonian, yielding similar results to those obtained from a plane-wave calculation using Wannier functions \cite{BenchmarkPaper}.

A large supercell was thus necessary for the Hamiltonian,
so that each of the four SWNT layers adjacent to the
boundaries $\mathpzc{H}_{C}^{{prin}}$, were within 0.1 eV of the Hamiltonian for the
respective leads $\mathpzc{H}_\alpha$, i.e. $\max |\mathpzc{H}_{C}^{{prin}} -\mathpzc{H}_\alpha| <$ 0.1 eV.  In this way the electronic structure at the
edges of the central region was ensured to be converged to that in the leads.

The Landauer-B\"{u}tticker conductance for a multi-terminal system can
be calculated from the Green's function of the central region,
$G_C$, according to the formula \cite{Meir,Thygesen2,BenchmarkPaper} 
\begin{equation}
G = \left. G_0
  \mathrm{Tr}[G_C\Gamma_{{in}}G_C^\dagger\Gamma_{{out}}]\right|_{\varepsilon = \varepsilon_F}, 
\end{equation}
where the trace runs over all localized basis functions in the central
region, and $G_0 \equiv {2}e^2/h$ is the quantum of conductance.  To describe the conductance at small bias for semiconducting
systems, the Fermi energy $\varepsilon_F$ should be taken as the
energy of the valence-band maximum $\varepsilon_{{VB}}$ or
conduction-band minimum $\varepsilon_{{CB}}$ for $p$-type and
$n$-type semiconductors, respectively. The central region Green's
function is calculated from  
\begin{equation}
G_C(\varepsilon) = \left[z \mathpzc{S}_C - \mathpzc{H}_C -
\sum_\alpha \Sigma_\alpha(\varepsilon)\right]^{-1},
\end{equation}
where $z = \varepsilon + i 0^+$, $\mathpzc{S}_C$ and $\mathpzc{H}_C$ are the overlap matrix and
Kohn-Sham Hamiltonian matrix of the central region in the localized
 basis, $\Sigma_{\alpha}$ is the self-energy of lead $\alpha$,  
\begin{equation}
\Sigma_\alpha(\varepsilon) = [z \mathpzc{S}_{C\alpha} -
  \mathpzc{H}_{C\alpha}] \left[z \mathpzc{S}_\alpha -
  \mathpzc{H}_\alpha\right]^{-1} [z \mathpzc{S}_{C\alpha}^\dagger
- \mathpzc{H}_{C\alpha}^\dagger],
\end{equation}
and the coupling elements between the central region and lead $\alpha$
for the overlap and Kohn-Sham Hamiltonian are $\mathpzc{S}_{C\alpha}$ and
$\mathpzc{H}_{C\alpha}$ respectively.

The coupling strengths of the input and output leads are then given by
 $\Gamma_{{in}/{out}} = i(\Sigma_{{in}/{out}} - \Sigma_{{in}/{out}}^\dagger)$.


In the following sections, we will analyze the transmission data for the clean TM doped SWNTs. To aid our understanding we will also consider the DOS of the system, which drives the behavior we see in the transmission. We will first analyze all the different metals in the divancancy II, then the same will be done with the divacancy I and the monovacancy.  Since we will concentrate on the DOS, it is important to understand the way the DOS and transmission T($\varepsilon$) are related. The peaks found in the DOS match with ``dips'' in the transmission, which are Fano anti-resonances \cite{Furst}. While narrow peaks in the DOS denote localized states, broad peaks denote strong bonding or hybridized states.

\subsection{Divacancy II}

\begin{figure}[!t]
\centering
\includegraphics[width=\textwidth]{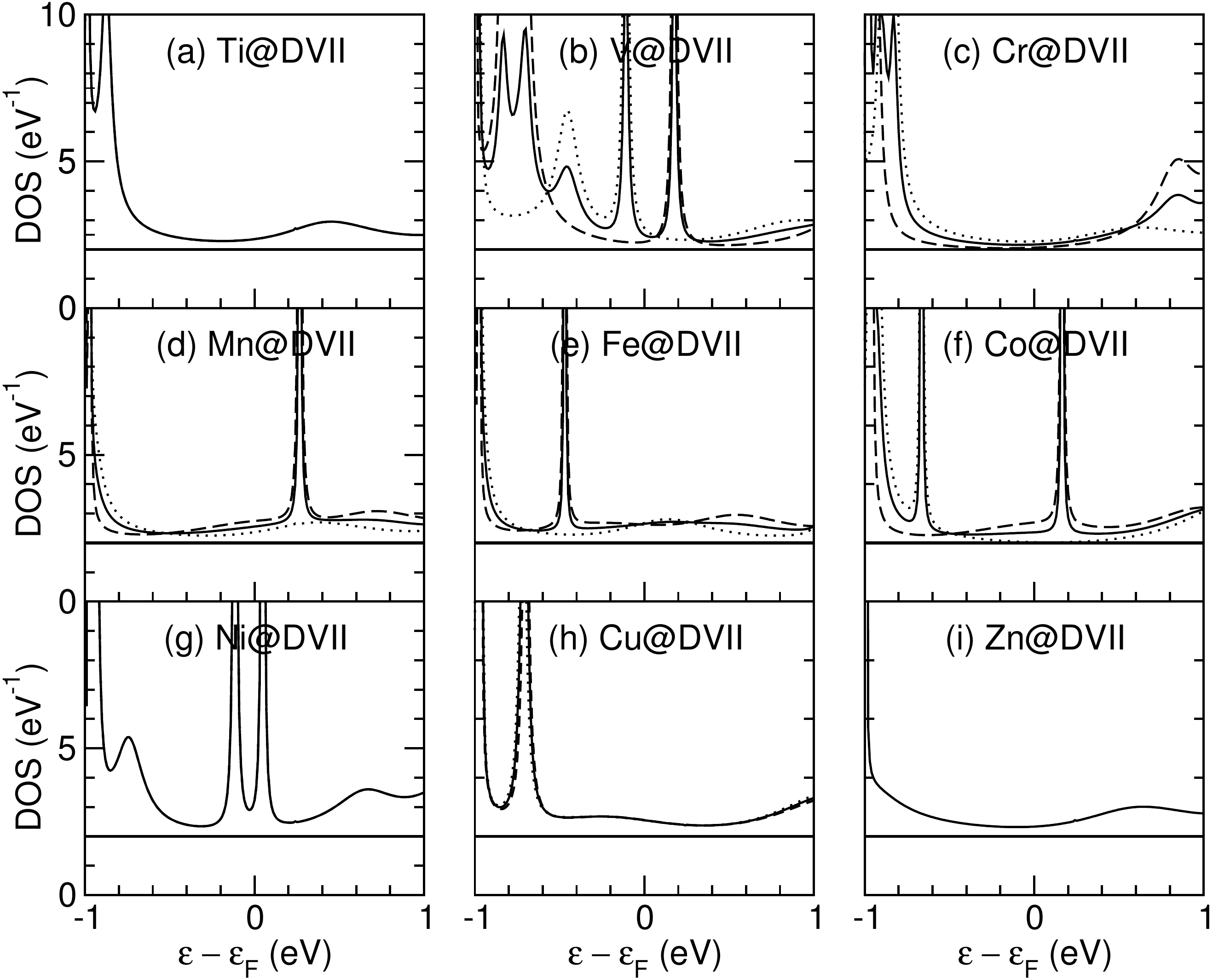}
\caption{Total density of states (DOS) in eV$^{-1}$ versus energy $\varepsilon$ in eV relative to the Fermi level $\varepsilon_F$ for a 3$d$ transition metal (TM) occupied divacancy II (DVII) in a (6,6) metallic armchair SWNT (TM@DVII).  Results for spin $\uparrow$ ($\cdot$ $\cdot$ $\cdot$ $\cdot$), spin $\downarrow$ (-- -- --), and spin averaged $\frac{\uparrow+\downarrow}{2}$ (------) are shown.  The constant DOS at 2 eV$^{-1}$ for a pristine (6,6) SWNT is provided for reference.}\label{DVII_Pure}
\end{figure}

\begin{figure}[!t]
\centering
\includegraphics[width=\textwidth]{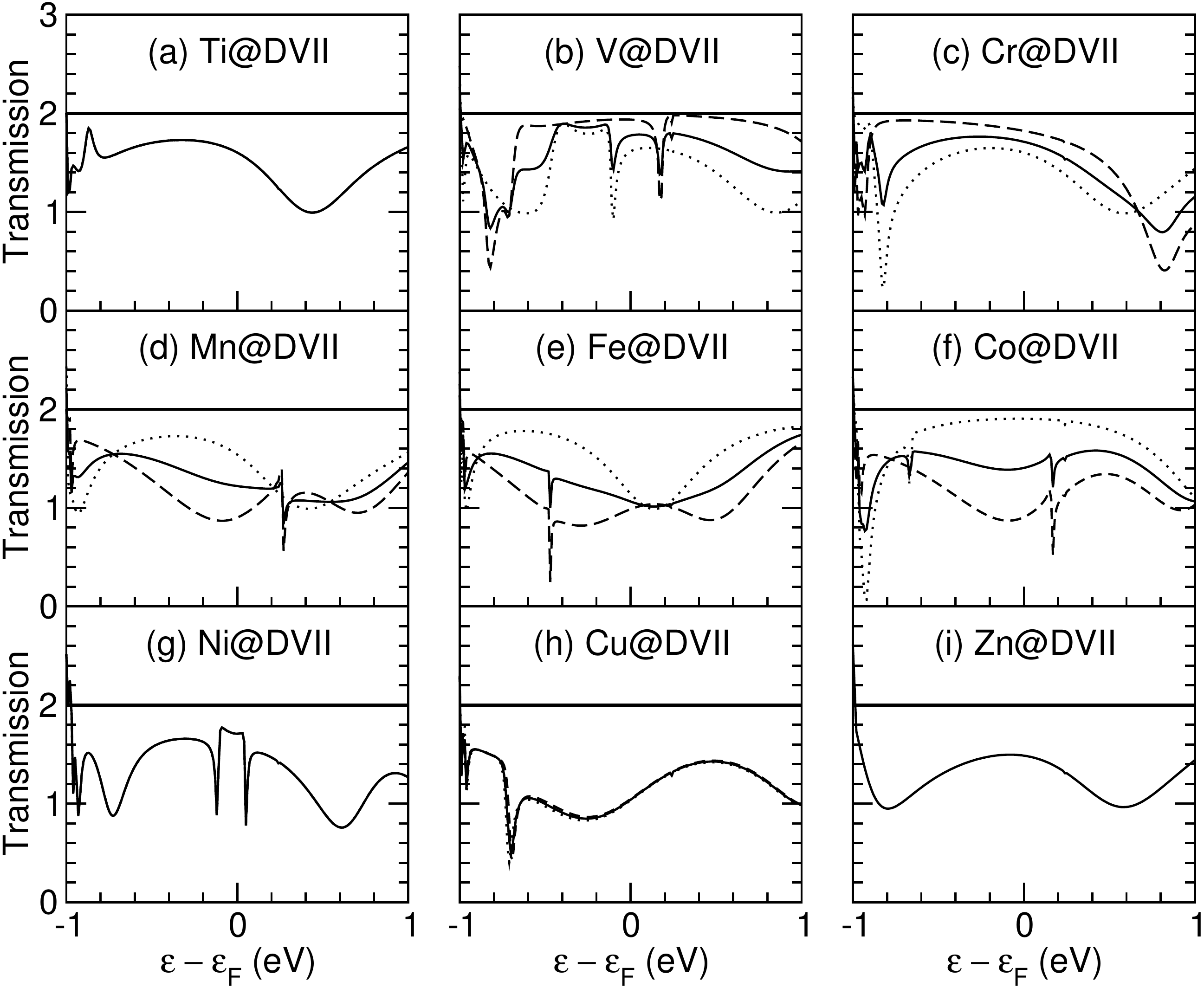}
\caption{Transmission versus energy $\varepsilon$ in eV relative to the Fermi level $\varepsilon_F$ through a 3$d$ transition metal (TM) occupied divacancy II (DVII) in a (6,6) metallic armchair SWNT (TM@DVII).  Results for spin $\uparrow$ ($\cdot$ $\cdot$ $\cdot$ $\cdot$), spin $\downarrow$ (-- -- --), and spin averaged $\frac{\uparrow+\downarrow}{2}$ (------) are shown.  The constant transmission of 2 for a pristine (6,6) SWNT is provided for reference.}\label{DVII_T}
\end{figure}

For a TM occupying a divacancy II (DVII) in a (6,6) SWNT (TM@DVII), shown schematically in Figure \ref{Stability}, there is an octahedral bonding of the TM atom, similar to the bonding in rutile metal oxides. Here, the TM bonding with the SWNT is through the two hybridized $\frac{d_{xz} \pm d_{yz}}{\sqrt{2}}$ $3d$ orbitals of the TM.  From Figure \ref{Stability} we see that all the TM--C bonds  are equally strained by the curvature of the nanotube.  This means that all TM--C bonds are equivalent.  To put it in other words, there is one type of four-fold degenerate TM--C bonding state.

The DOS and transmission T$(\varepsilon)$ through a pure TM occupied divacancy (TM@ DVII) are shown in Figures \ref{DVII_Pure} and \ref{DVII_T}, respectively. As mentioned, if we compare the DOS and T$(\varepsilon)$ shown in Figures \ref{DVII_Pure} and \ref{DVII_T}, we find  peaks and dips, respectively, relative to the flat DOS and T$(\varepsilon)$ of a pristine (6,6) SWNT. 

For Ti@DVII there are four occupied C $\pi$---Ti$\frac{d_{xz} \pm d_{yz}}{\sqrt{2}}$ bonds. These bonds are more than 1 eV below the Fermi level, so that they are not relevant for this study. There are also present completely unoccupied $d_{x^2-y^2}$ orbitals and anti-bonding Ti $\frac{d_{xz} \pm d_{yz}}{\sqrt{2}}$ orbitals. However, only the $t_{2g}$ $d$ orbitals of the TMs are important in the $-1$ to 1 eV range. These $d$ orbitals are what determines the binding and transmission for the TM@DVII system. 

In the case of V@DVII, there is one extra electron in the $d_{x^2-y^2}$ orbital, which is a non-bonding orbital. This extra electron yields a spin splitting in the DOS, as seen in Figure \ref{DVII_Pure}. The $d_{x^2-y^2}$ orbital is, by symmetry, non-bonding and it is localized on the V atom.

For Cr@DVII there are two extra electrons in the $d_{x^2-y^2}$ orbital, instead of only one. The states are shifted down in energy and there is no state around the Fermi level, as seen in Figure \ref{DVII_Pure}.

With Mn@DVII, electrons begin occupying the $t_{2g}$ anti-bonding orbitals. There are then localized spin states above the Fermi level. These spin unpaired $d$ states, which are localized on the TM atom, move down in energy as they are filled with more $d$-electrons. This is what we see in Figure \ref{DVII_Pure} for Fe@DVII and Co@DVII.

For Ni@DVII, these localized states become spin-paired, with one of them being completely occupied, and the other one completely empty. When we add another electron to the $d$ band, moving to Cu@DVII, both states become fully occupied,  localized on the Cu atom, and spin-paired.

Finally, for Zn@DVII, the $d$-orbitals are now completely filled. The DOS closely resembles that of Ti@DVII, and the states are spin-paired, as shown in Figure \ref{DVII_Pure}.

\subsection{Divacancy I}

\begin{figure}[!t]
\centering
\includegraphics[width=\textwidth]{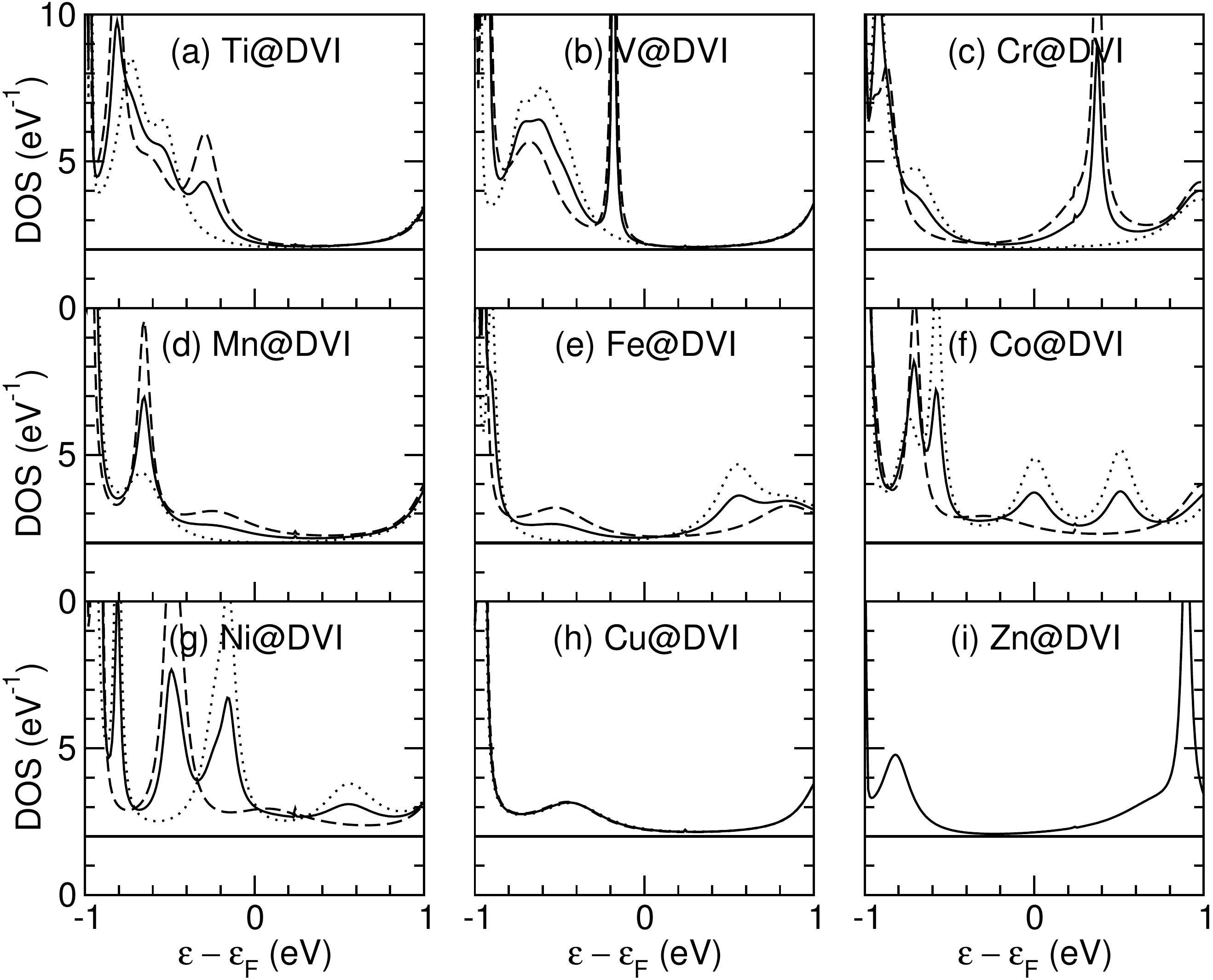}
\caption{Total density of states (DOS) in eV$^{-1}$ versus energy $\varepsilon$ in eV relative to the Fermi level $\varepsilon_F$ for a 3$d$ transition metal (TM) occupied divacancy I (DVI) in a (6,6) metallic armchair SWNT (TM@DVI).  Results for spin $\uparrow$ ($\cdot$ $\cdot$ $\cdot$ $\cdot$), spin $\downarrow$ (-- -- --), and spin averaged $\frac{\uparrow+\downarrow}{2}$ (------) are shown.  The constant DOS at 2 eV$^{-1}$ for a pristine (6,6) SWNT is provided for reference.}\label{DVI_Pure}
\end{figure}

\begin{figure}[!t]
\centering
\includegraphics[width=\textwidth]{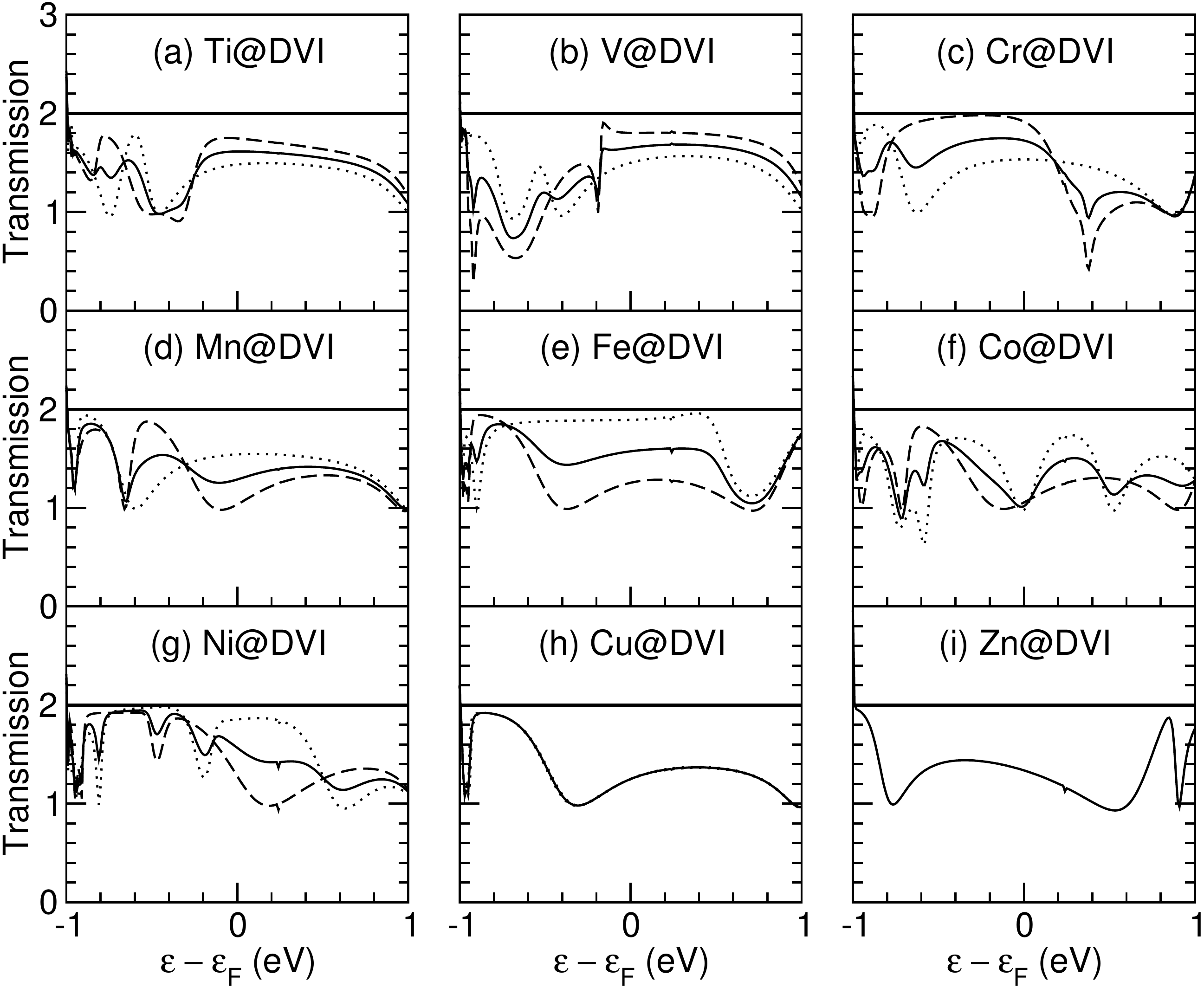}
\caption{Transmission versus energy $\varepsilon$ in eV relative to the Fermi level $\varepsilon_F$ through a 3$d$ transition metal (TM) occupied divacancy I (DVI) in a (6,6) metallic armchair SWNT (TM@DVI).  Results for spin $\uparrow$ ($\cdot$ $\cdot$ $\cdot$ $\cdot$), spin $\downarrow$ (-- -- --), and spin averaged $\frac{\uparrow+\downarrow}{2}$ (------) are shown.  The constant transmission of 2 for a pristine (6,6) SWNT is provided for reference.}\label{DVI_T}
\end{figure}

For a TM occupying a divacancy I (DVI) in a (6,6) SWNT (TM@DVI), shown schematically in Figure \ref{Stability}, the states are no longer as localized on the metal atom as for the TM@DVII, and are more strongly hybridized with the C $\pi$ states. Due to the geometric configuration of TM@DVI, the angle of two of the bonds with the axis of the nanotube is much smaller than the angle of the other two bonds. We will call these two different bonds near-axis and off-axis, respectively. This difference in the orientation produces a difference in the strain the bonds will have to withstand. 

The off-axis bonds will undergo more stress due the SWNT's curvature, compared to the near-axis bonds. This means that there are two types of degenerate TM--C bonding states.  This degeneration results in a larger number of broader peaks and dips, compared to those obtained for a TM@DVII, as shown in Figures \ref{DVI_Pure} and \ref{DVI_T}, respectively

\begin{figure}[!th]
\centering
\includegraphics[width=\textwidth]{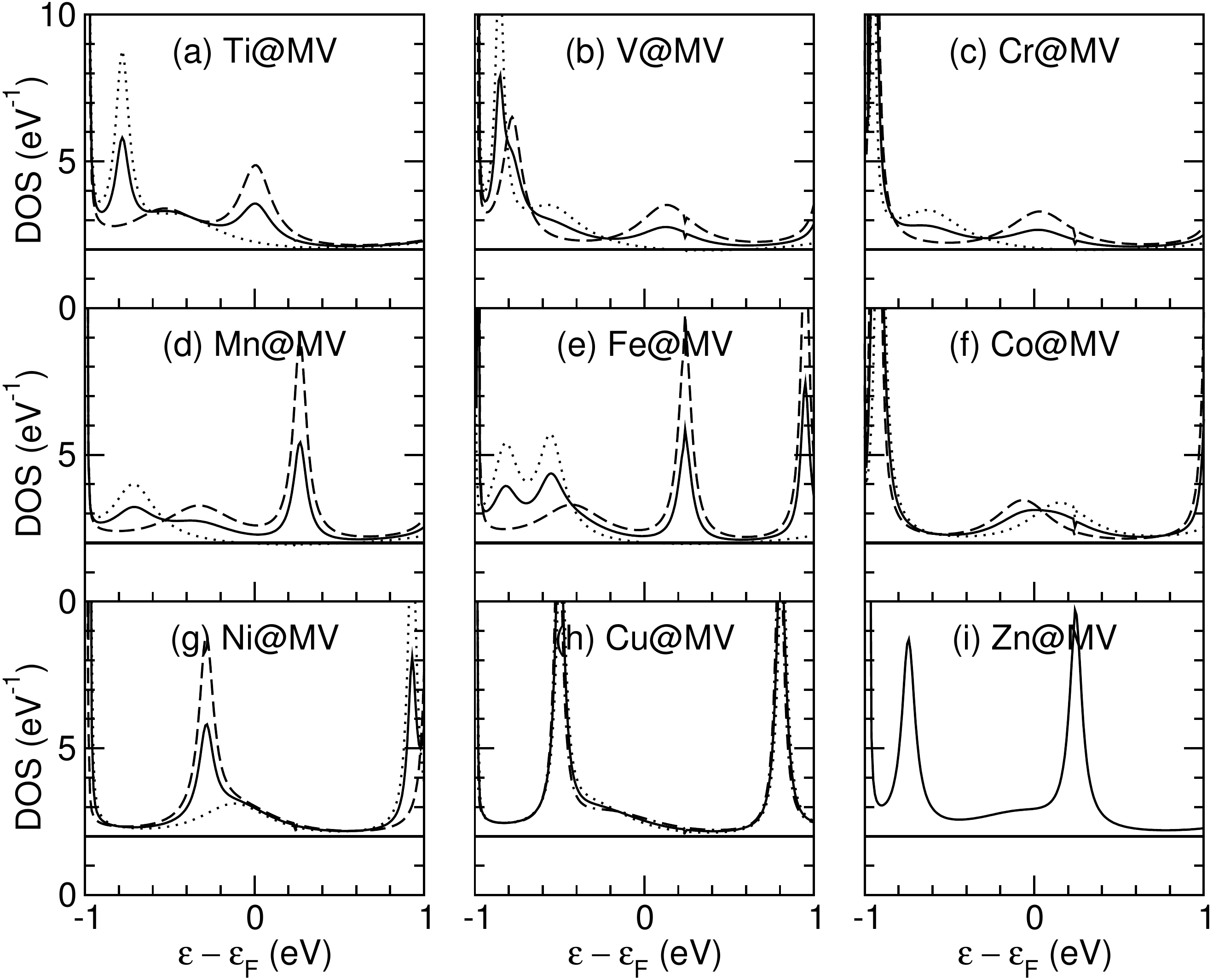}
\caption{Total density of states (DOS) in eV$^{-1}$ versus energy $\varepsilon$ in eV relative to the Fermi level $\varepsilon_F$ for a 3$d$ transition metal (TM) occupied monovacancy (MV) in a (6,6) metallic armchair SWNT (TM@MV).  Results for spin $\uparrow$ ($\cdot$ $\cdot$ $\cdot$ $\cdot$), spin $\downarrow$ (-- -- --), and spin averaged $\frac{\uparrow+\downarrow}{2}$ (------) are shown.  The constant DOS at 2 eV$^{-1}$ for a pristine (6,6) SWNT is provided for reference.}\label{MV_DOS}
\end{figure}

\begin{figure}[!t]
\centering
\includegraphics[width=\textwidth]{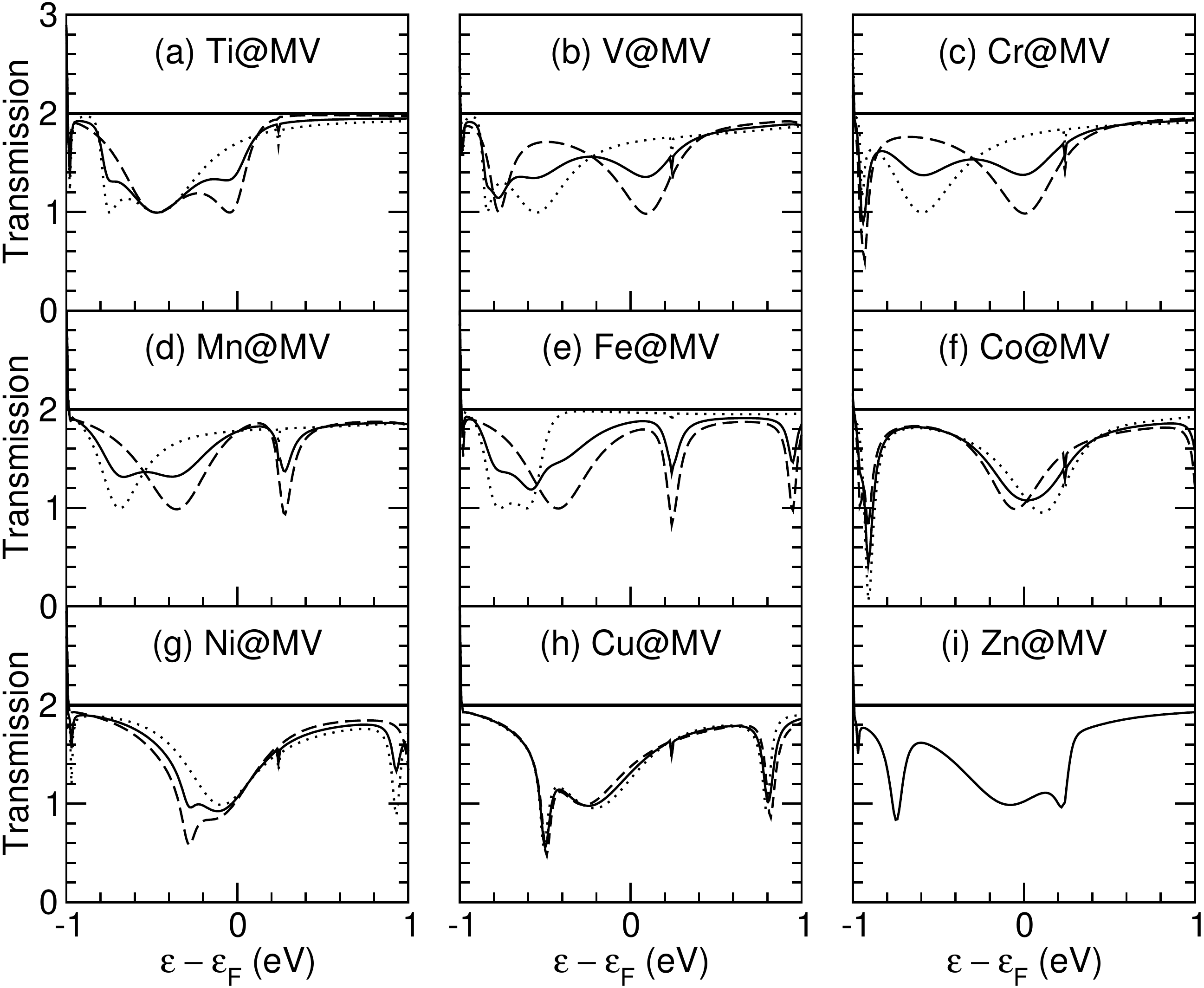}
\caption{Transmission versus energy $\varepsilon$ in eV relative to the Fermi level $\varepsilon_F$ through a 3$d$ transition metal (TM) occupied monovacancy (MV) in a (6,6) metallic armchair SWNT (TM@MV).  Results for spin $\uparrow$ ($\cdot$ $\cdot$ $\cdot$ $\cdot$), spin $\downarrow$ (-- -- --), and spin averaged $\frac{\uparrow+\downarrow}{2}$ (------) are shown.  The constant transmission of 2 for a pristine (6,6) SWNT is provided for reference.}\label{MV_T}
\end{figure}

Even with these differences it is still possible to see in Figure \ref{DVI_Pure} the filling of states, and their movement to lower energy, exactly like in the case of the TM@DVII. It is very likely that the differences in the strain  contribute to the splitting of the spin degeneracy for Ti@DVI and Ni@DVI.

The Ti@DVI $d$ orbitals are half-filled through bonding with the C $\pi$ orbitals, as is the 4$s$ level. The Cu@DVI $d$ orbitals are filled by taking an electron from the 4$s$ level. So, in the case of Cu@DVI, bonding occurs through the half-occupied 4$s$ orbital. All of the Zn $d$ orbitals are completely filled, as is the 4$s$ level. For these three systems, the DOS shown in Figure \ref{DVI_Pure} is very similar to that shown in Figure \ref{DVII_Pure} for the TM@DVII case. However, for the Ti@DVI DOS there is a noticeable difference, most probably due to the differences in strain for the two types of C--TM bonds (near-axis and off-axis).

\subsection{Monovacancy}

\begin{figure}[!t]
\centering
\includegraphics[width=\textwidth]{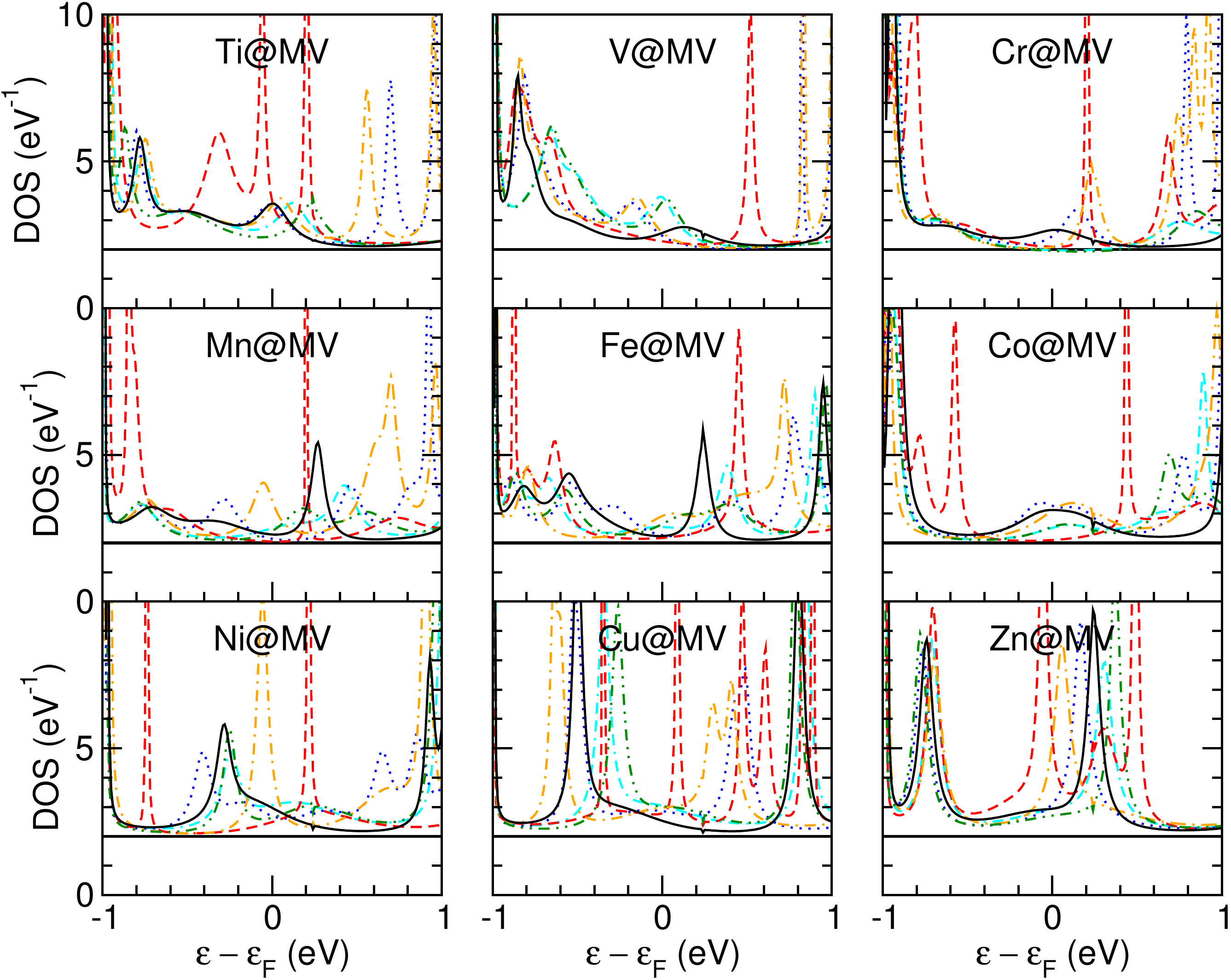}
\caption{Total density of states (DOS) in eV$^{-1}$ versus energy $\varepsilon$ in eV relative to the Fermi level $\varepsilon_F$ for a 3$d$ transition metal (TM) occupied monovacancy (MV) in a (6,6) metallic armchair SWNT (TM@MV).  Results for the clean TM@MV ({\bf{---------}}) and with adsorbed N$_2$ ({\bf{\color{blue}{$\cdot$ $\cdot$ $\cdot$ $\cdot$ $\cdot$ $\cdot$}}}), O$_2$ ({\bf{\color{red}{--~--~--~--}}}), H$_2$O ({\bf{\color{cyan}{--~--~$\cdot$ --~--}}}), CO ({\bf{\color{Orange}{--~$\cdot$ --~$\cdot$ --}}}), and NH$_3$ ({\bf{\color{Green}{--~$\cdot$ $\cdot$ --~$\cdot$ $\cdot$}}}) are shown.  The constant DOS at 2 eV$^{-1}$ for a pristine (6,6) SWNT is provided for reference.}\label{MV_DOS_ALL}
\end{figure}

\begin{figure}[!t]
\centering
\includegraphics[width=\textwidth]{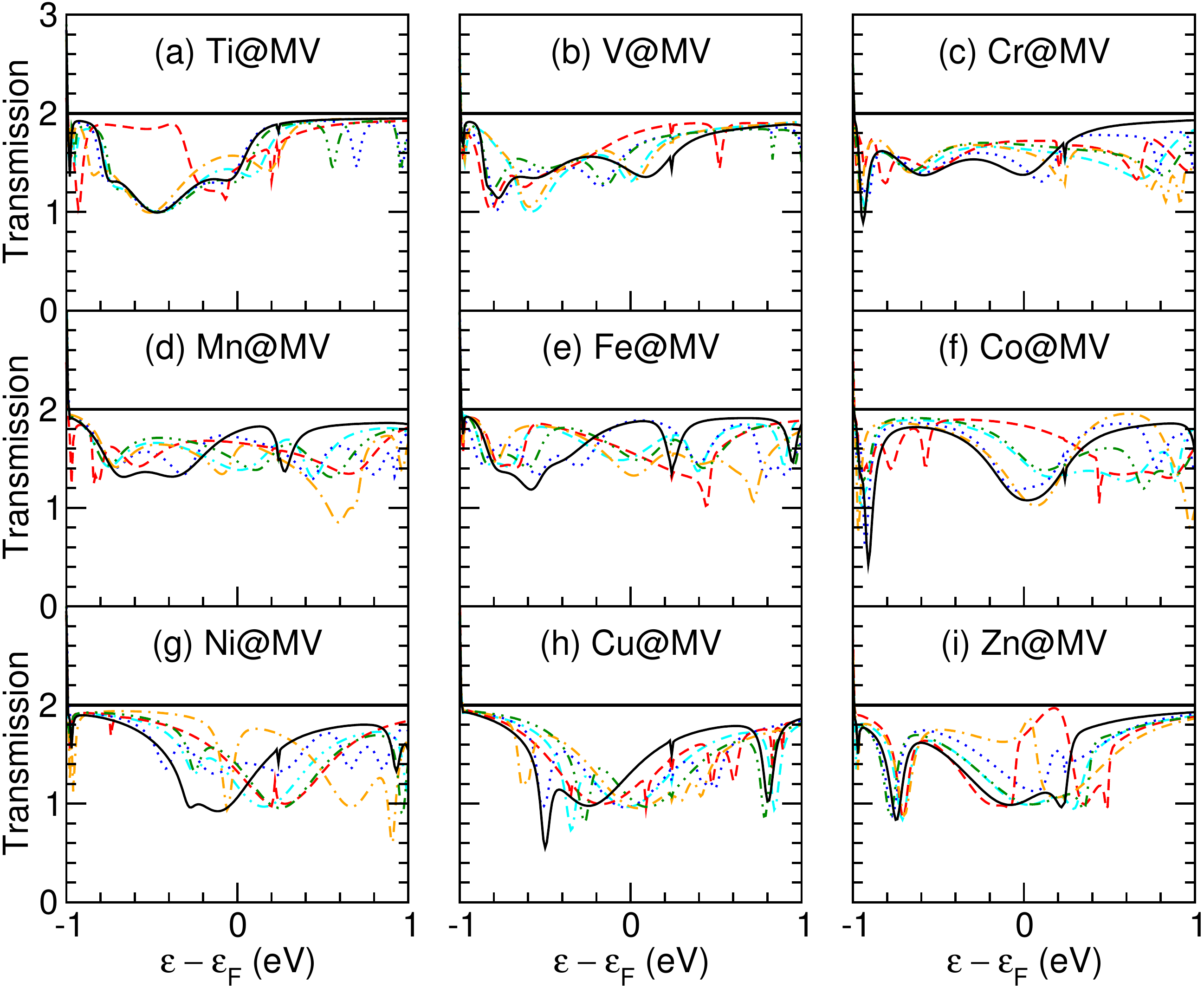}
\caption{Transmission versus energy $\varepsilon$ in eV relative to the Fermi level $\varepsilon_F$ through a 3$d$ transition metal (TM) occupied monovacancy (MV) in a (6,6) metallic armchair SWNT (TM@MV).  Results for the clean TM@MV ({\bf{---------}}) and with adsorbed N$_2$ ({\bf{\color{blue}{$\cdot$ $\cdot$ $\cdot$ $\cdot$ $\cdot$ $\cdot$}}}), O$_2$ ({\bf{\color{red}{--~--~--~--}}}), H$_2$O ({\bf{\color{cyan}{--~--~$\cdot$ --~--}}}), CO ({\bf{\color{Orange}{--~$\cdot$ --~$\cdot$ --}}}), and NH$_3$ ({\bf{\color{Green}{--~$\cdot$ $\cdot$ --~$\cdot$ $\cdot$}}}) are shown.  The constant transmission of 2 for a pristine (6,6) SWNT is provided for reference.}\label{MV_AT}
\end{figure}

%
\begin{figure}[!t]
\centering
\includegraphics[width=\textwidth]{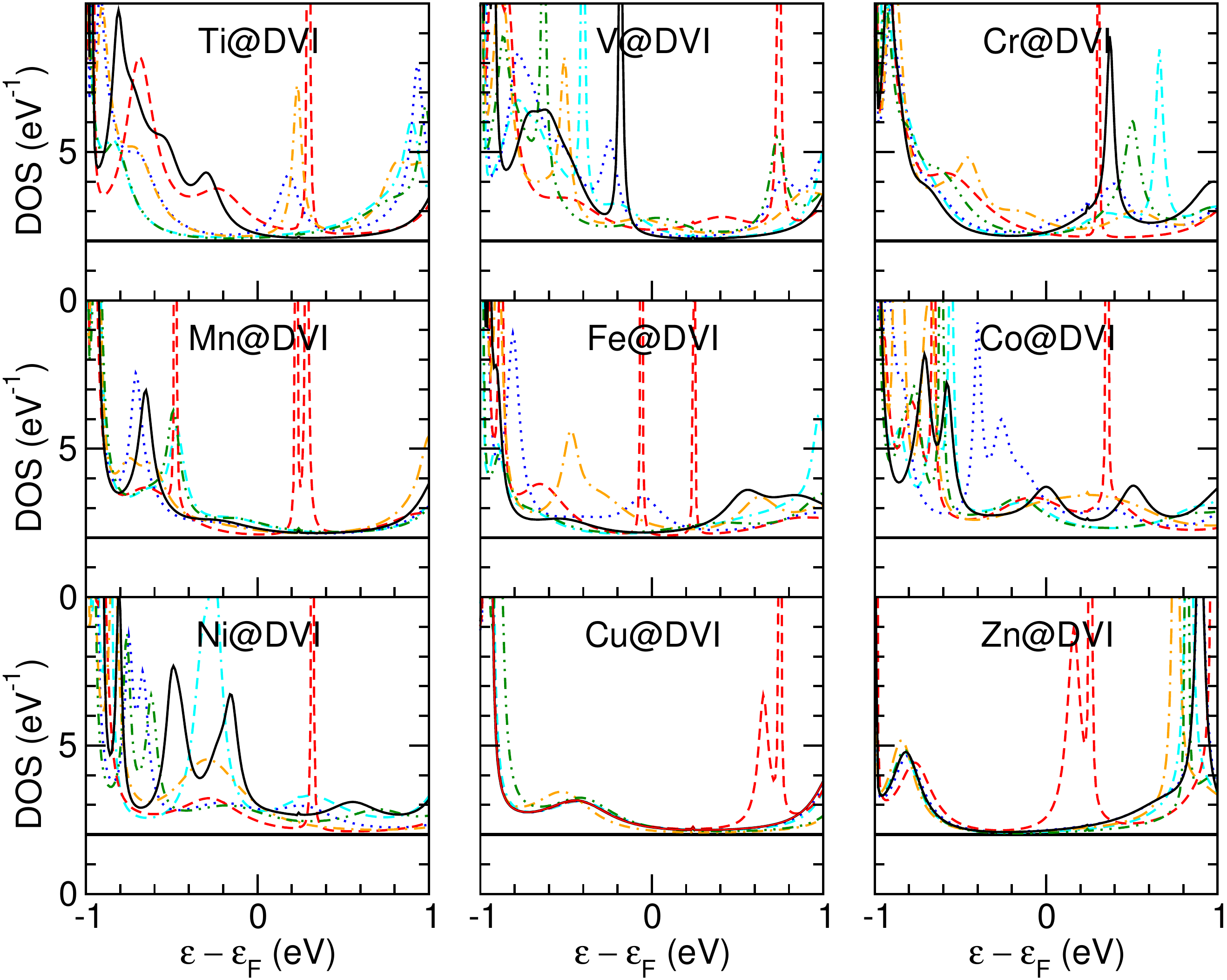}
\caption{Total density of states (DOS) in eV$^{-1}$ versus energy $\varepsilon$ in eV relative to the Fermi level $\varepsilon_F$ for a 3$d$ transition metal (TM) occupied divacancy I (DVI) in a (6,6) metallic armchair SWNT (TM@DVI).  Results for the clean TM@MV ({\bf{---------}}) and with adsorbed N$_2$ ({\bf{\color{blue}{$\cdot$ $\cdot$ $\cdot$ $\cdot$ $\cdot$ $\cdot$}}}), O$_2$ ({\bf{\color{red}{--~--~--~--}}}), H$_2$O ({\bf{\color{cyan}{--~--~$\cdot$ --~--}}}), CO ({\bf{\color{Orange}{--~$\cdot$ --~$\cdot$ --}}}), and NH$_3$ ({\bf{\color{Green}{--~$\cdot$ $\cdot$ --~$\cdot$ $\cdot$}}}) are shown.  The constant DOS at 2 eV$^{-1}$ for a pristine (6,6) SWNT is provided for reference.}\label{DVI_DOS_ALL}
\end{figure}

\begin{figure}[!t]
\centering
\includegraphics[width=\textwidth]{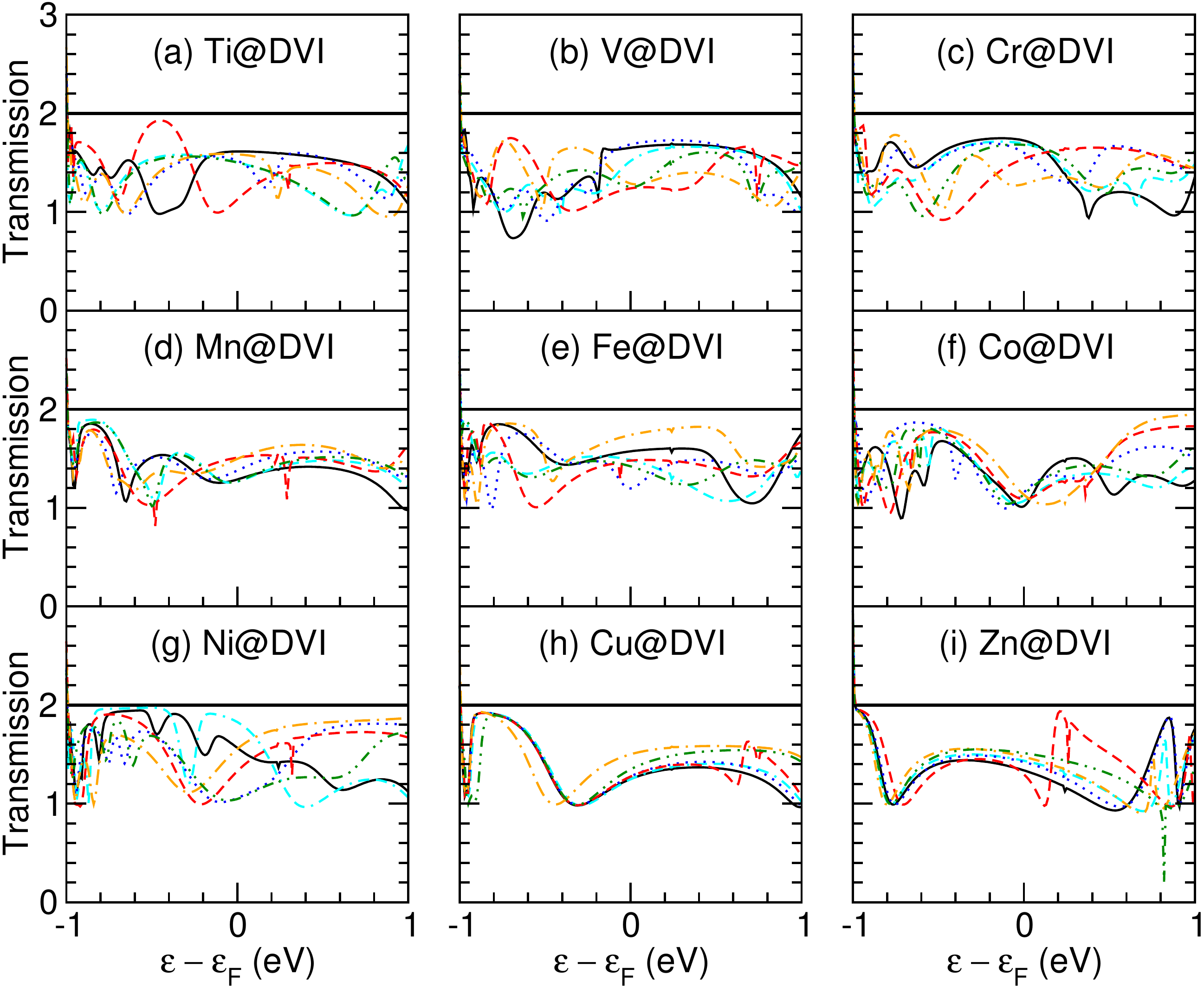}
\caption{Transmission versus energy $\varepsilon$ in eV relative to the Fermi level $\varepsilon_F$ through a 3$d$ transition metal (TM) occupied divacancy I (DVI) in a (6,6) metallic armchair SWNT (TM@DVI).  Results for the clean TM@DVI ({\bf{---------}}) and with adsorbed N$_2$ ({\bf{\color{blue}{$\cdot$ $\cdot$ $\cdot$ $\cdot$ $\cdot$ $\cdot$}}}), O$_2$ ({\bf{\color{red}{--~--~--~--}}}), H$_2$O ({\bf{\color{cyan}{--~--~$\cdot$ --~--}}}), CO ({\bf{\color{Orange}{--~$\cdot$ --~$\cdot$ --}}}), and NH$_3$ ({\bf{\color{Green}{--~$\cdot$ $\cdot$ --~$\cdot$ $\cdot$}}}) are shown.  The constant transmission of 2 for a pristine (6,6) SWNT is provided for reference.}\label{DVI_AT}
\end{figure}

%
\begin{figure}[!t]
\centering
\includegraphics[width=\textwidth]{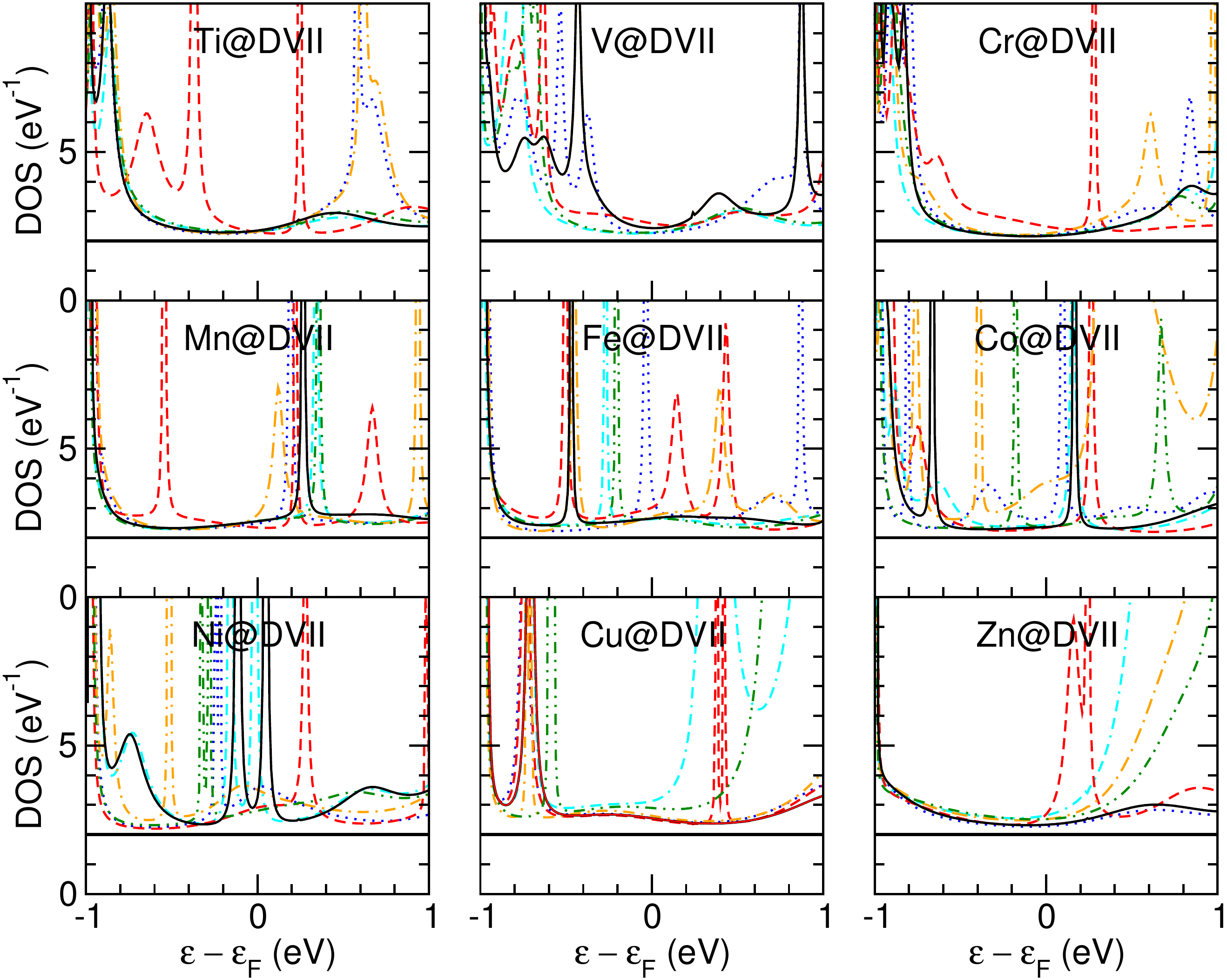}
\caption{Total density of states (DOS) in eV$^{-1}$ versus energy $\varepsilon$ in eV relative to the Fermi level $\varepsilon_F$ for a 3$d$ transition metal (TM) occupied divacancy II (DVII) in a (6,6) metallic armchair SWNT (TM@DVII).  Results for the clean TM@MV ({\bf{---------}}) and with adsorbed N$_2$ ({\bf{\color{blue}{$\cdot$ $\cdot$ $\cdot$ $\cdot$ $\cdot$ $\cdot$}}}), O$_2$ ({\bf{\color{red}{--~--~--~--}}}), H$_2$O ({\bf{\color{cyan}{--~--~$\cdot$ --~--}}}), CO ({\bf{\color{Orange}{--~$\cdot$ --~$\cdot$ --}}}), and NH$_3$ ({\bf{\color{Green}{--~$\cdot$ $\cdot$ --~$\cdot$ $\cdot$}}}) are shown.  The constant DOS at 2 eV$^{-1}$ for a pristine (6,6) SWNT is provided for reference.}\label{DVII_DOS_ALL}
\end{figure}

\begin{figure}[!t]
\centering
\includegraphics[width=\textwidth]{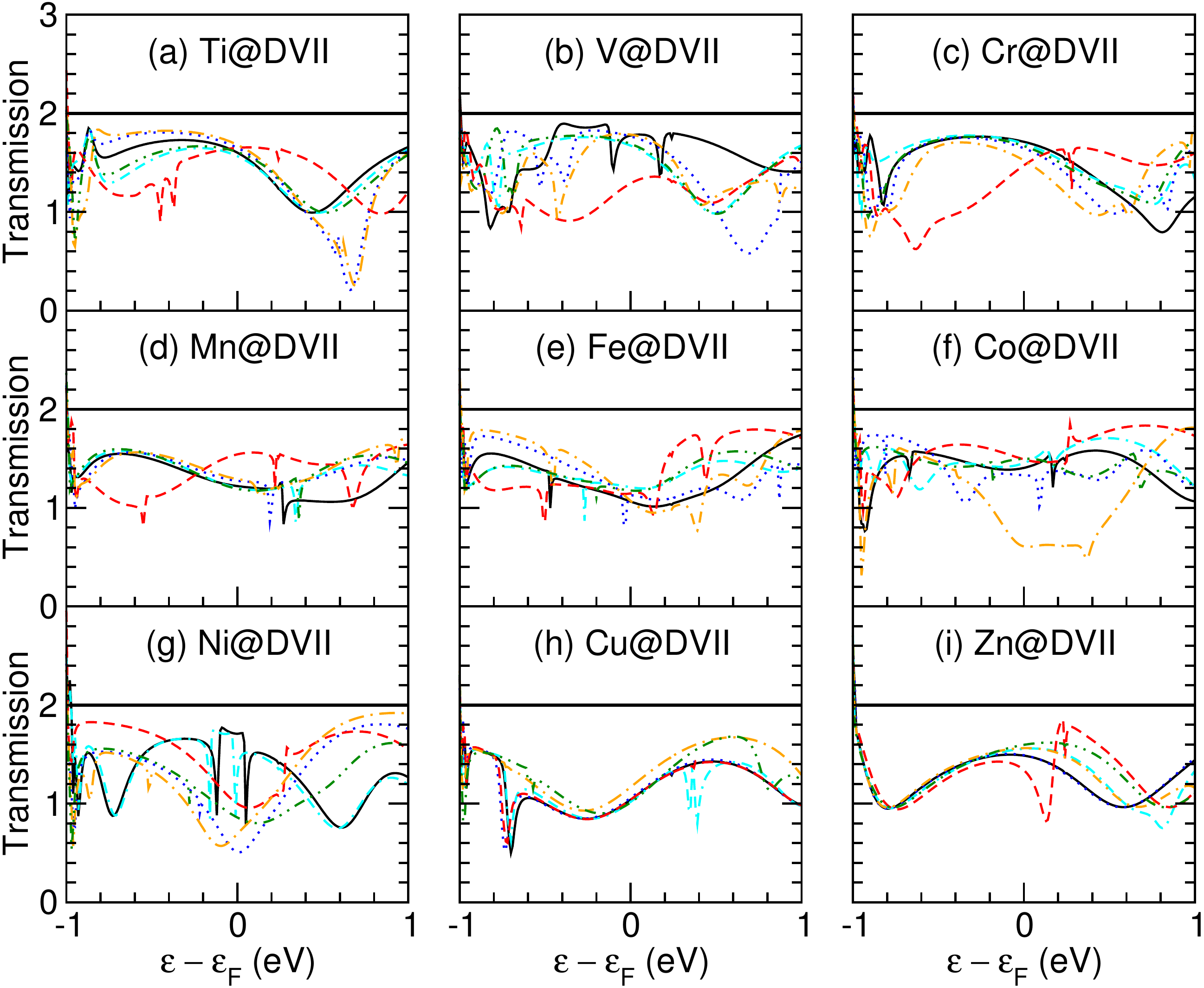}
\caption{Transmission versus energy $\varepsilon$ in eV relative to the Fermi level $\varepsilon_F$ through a 3$d$ transition metal (TM) occupied divacancy II (DVII) in a (6,6) metallic armchair SWNT (TM@DVII).  Results for the clean TM@DVII ({\bf{---------}}) and with adsorbed N$_2$ ({\bf{\color{blue}{$\cdot$ $\cdot$ $\cdot$ $\cdot$ $\cdot$ $\cdot$}}}), O$_2$ ({\bf{\color{red}{--~--~--~--}}}), H$_2$O ({\bf{\color{cyan}{--~--~$\cdot$ --~--}}}), CO ({\bf{\color{Orange}{--~$\cdot$ --~$\cdot$ --}}}), and NH$_3$ ({\bf{\color{Green}{--~$\cdot$ $\cdot$ --~$\cdot$ $\cdot$}}}) are shown.  The constant transmission of 2 for a pristine (6,6) SWNT is provided for reference.}\label{DVII_AT}
\end{figure}

For a TM occupying a monovacancy (MV) in a (6,6) SWNT (TM@MV), shown schematically in Figure \ref{Stability}, the TM atom has three bonds with three C atoms, two of the bonds being shorter, stronger, and near-axis, and one longer, weaker, and perpendicular to the nanotube axis. Not only is this bond weaker, but it is also the one that withstands the largest strain.  From Figure \ref{MV_DOS} we see that for Ti@MV and Ni@MV we have spin splitting, which may be attributable to the strain in the weaker C--TM bond. Figure \ref{MV_DOS} also shows that for the monovacancy the DOS shifts down in energy as the $d$ orbitals on the TM atom are filled.  The differences between the bond lengths and strengths of the three C--TM bonds yields several broad peaks in the DOS, as seen in Figure \ref{MV_DOS}, and likewise broad dips in the transmission, as seen in Figure \ref{MV_T}.

As shown in Figure \ref{MV_DOS}, for Ti@MV, V@MV, and Cr@MV, we see a broad spin-unpaired peak centered near the Fermi level. This peak is shifted down in energy for Mn@MV, which has five $d$ electrons. At this point a narrower peak, for a $d$-state more localized on the TM atom enters the energy window and it is positioned at round 0.2 eV.  This state broadens for Co@MV where it is partly occupied, and nearly spin-paired.  For Ni@MV this state goes further down in energy relative to the Fermi level and becomes spin paired for Cu@MV and Zn@MV.

\subsection{Target and Background Molecules}

To estimate the effect of adsorbates on the electrical conductance of doped SWNTs, we first consider the change in conductance when a single molecule is adsorbed on a metal site of an otherwise pristine SWNT.  In Figs.~\ref{MV_DOS_ALL}, \ref{DVI_DOS_ALL}, and \ref{DVII_DOS_ALL} we show the calculated DOS for a TM occupied monovacancy, divacancy I, and divacancy II, respectively, with and without an adsorbed molecule $X$. These may be compared with Figs.~\ref{MV_AT}, \ref{DVI_AT}, and \ref{DVII_AT} where we show the calculated transmission probability T for a TM occupied monovacancy, divacancy I, and divacancy II, respectively, with and without an adsorbed molecule $X$.   

In contrast to the adsorption energies, there are no clear trends in the conductances. The sensitivity of the conductance is perhaps most clearly demonstrated by the absence of correlation between different types of vacancies. 

Close to the Fermi level, the conductance of a perfect armchair SWNT equals 2$G_0$.  The presence of the metal dopant leads to several dips in the transmission function known as Fano anti-resonances  \cite{Furst}. The position and shape of these dips depends on the $d$-levels of the TM atom, the character of its bonding to the SWNT, and is further affected by the presence of the adsorbate molecule. The coupling of all these factors is very complex and makes it difficult to estimate or rationalize the value of the conductance. For the spin-polarized cases, we use the spin-averaged conductances, i.e. $G=(G_{\uparrow}+G_{\downarrow})/2$.

Cu@DVI binds all molecules rather weakly, so any changes in the DOS
from the clean Cu@DVI structure between $-1$ and 1 eV are
attributable to states completely localized on the molecule.  Here we
only see two states localized on the O$_2$ molecule between 0.6 and 0.8
eV above the Fermi level.
For CO+Cu@DVI we see a shifting of the DOS, and hybridization with the
Cu@DVI states.  In particular, the broad peak below $\varepsilon_F$ in
the Cu@DVI DOS, which is binding with the C $\pi$ states,  is shifted
down in energy by about 0.1 eV when CO is adsorbed on the Cu@DVI site.

For O$_2$, its ground state has spin 2, with one electron occupying a
$p_z$---$ p_z$ anti-bonding $\pi^*$ orbital, and the other electron
occupying a $p_y$---$ p_y$ anti-bonding $\pi^*$ orbital
\cite{cnt_networks}.  For O$_2$ on Ti@DVI, V@DVI, and Cr@DVI, the $p_z$---$ p_z$ anti-bonding $\pi^*$ orbital hybridizes with the transition
metal atom's $d_{yz}$ state, leaving the spin down component of the
other $p_y$---$ p_y$ anti-bonding $\pi^*$ state on the molecule
unoccupied.  This state is visible as a sharp peak above the Fermi
level.  By symmetry the $p_y$---$ p_y$ anti-bonding O$_2$ $\pi^*$
state cannot bond to the TM@DVI system.

It should be noted that O$_2$ is the only molecule with molecular states in the $-1$ to +1 eV range relative to the (6,6) SWNT Fermi level. The O$_2$ $p_y$---$p_y$ anti-bonding $\pi^*$ states by symmetry are forbidden from bonding with the TM@DVI and TM@DVII systems. On the other hand, the O$_2$ $p_z$---$ p_z$ anti-bonding $\pi^*$ state overlaps with the TM $d_{yz}$ state. For Cu and Zn, as the $d$ orbitals are filled, we find the O$_2$ anti-bonding $\pi^*$ states are both spin unpaired, and exhibit little hybridization with the TM@VC site.

Overall, the TM $d$ states are shifted up and down in
energy when a molecule is adsorbed, based on both charge transfer, and
the symmetry of the molecule's orbitals.  This leads to a rich
response in the DOS of the TM@VC system upon adsorption of a molecule.
The main conclusion to be made is that the DOS and thus the transmission through the
TM@VC systems are quite sensitive to the adsorption of a molecule, and
give a strong response. This makes the resistance through such active
sites an excellent sensing property.

\section{Sensing Property}

\begin{figure}[!th]
\centering
\includegraphics[scale=0.5]{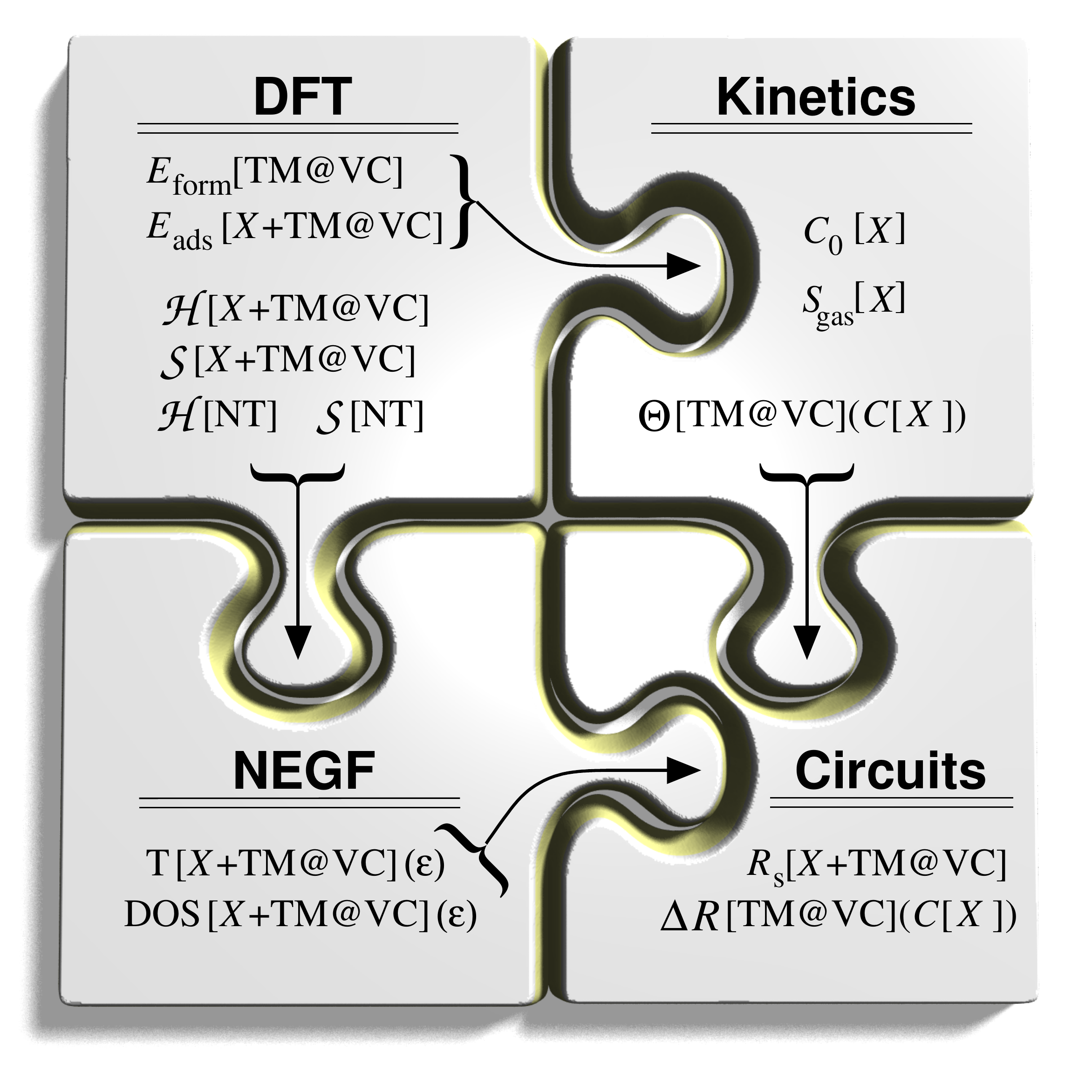}
\caption{The coverage $\Theta$ and transmission probabilities T$(\varepsilon)$ are used within simple circuit theory to estimate the scattering resistance $R_s$ for species $X$ on an active site ($X$+TM@VC), and hence derive the overall change in resistance $\Delta R$ for a particular active site (TM@VC) as a function of the target molecule's concentration $C[X]$.}\label{Circuits}
\end{figure}

The final ``piece of the puzzle'' for our model of a chemical nanosensor is the description of the sensing property, i.e. the change in resistance upon adsorption of a molecule under atmospheric conditions, \emph{via} basic circuit theory, as depicted in Figure \ref{Circuits}.  At this point we bring together the transmission T$(\varepsilon)$ calculated from the NEGF method and the coverage of active sites $\Theta[X]$ from the kinetic model to describe the change in resistance $\Delta R$ through the doped SWNT as a function of the target molecule's concentration $C[X]$.


\begin{table}
\caption{Conductance $G[X]$ in $G_0 \equiv 2e^2/h$ for a $3d$ transition metal occupied monovacancy (TM@MV) with N$_2$, O$_2$, H$_2$O, CO, NH$_3$, and H$_2$S adsorbed.  The conductance of the clean TM@MV is provided for comparison.}\label{TM@MV_G}
\begin{tabular}{l|ccccccccc}
 & Ti & V & Cr & Mn & Fe & Co & Ni & Cu & Zn\\\hline
N$_2$ & 1.44 & 1.63 & 1.44 & 1.72 & 1.88 & 1.19 & 1.40 & 0.97 & 1.16\\
O$_2$ & 1.44 & 1.78 & 1.72 & 1.66 & 1.56 & 1.83 & 1.33 & 1.13 & 1.69\\
H$_2$O & 1.42 & 1.31 & 1.64 & 1.39 & 1.50 & 1.47 & 1.18 & 0.97 & 0.99\\
CO & 1.31 & 1.61 & 1.61 & 1.59 & 1.33 & 1.09 & 1.58 & 0.97 & 1.79\\
NH$_3$ & 1.57 & 1.37 & 1.68 & 1.47 & 1.51 & 1.50 & 1.39 & 1.05 & 1.01\\
H$_2$S & 1.40 & 1.51 & 1.64 & 1.51 & 1.50 & 1.43 & 1.32 & 0.98 & 1.13\\
Clean & 1.40 & 1.41 & 1.38 & 1.77 & 1.87 & 1.08 & 1.06 & 1.31 & 1.02\\
\end{tabular}
\end{table}

\begin{table}
\caption{Conductance $G[X]$ in $G_0 \equiv 2e^2/h$ for a $3d$ transition metal occupied divacancy I (TM@DVI) with N$_2$, O$_2$, H$_2$O, CO, NH$_3$, and H$_2$S adsorbed.  The conductance of the clean TM@DVI is provided for comparison.}\label{TM@DVI_G}
\begin{tabular}{l|ccccccccc}
 & Ti & V & Cr & Mn & Fe & Co & Ni & Cu & Zn\\\hline
N$_2$ & 1.52 & 1.69 & 1.60 & 1.38 & 1.20 & 1.16 & 1.06 & 1.28 & 1.42\\
O$_2$ & 1.13 & 1.24 & 1.54 & 1.51 & 1.48 & 1.09 & 1.33 & 1.28 & 1.30\\
H$_2$O & 1.50 & 1.52 & 1.69 & 1.27 & 1.49 & 1.07 & 1.83 & 1.25 & 1.44\\
CO & 1.58 & 1.29 & 1.28 & 1.50 & 1.74 & 1.13 & 1.56 & 1.50 & 1.48\\
NH$_3$ & 1.50 & 1.25 & 1.68 & 1.27 & 1.42 & 1.14 & 1.05 & 1.35 & 1.51\\
H$_2$S & 1.51 & 1.59 & 1.69 & 1.31 & 1.43 & 1.11 & 1.03 & 1.29 & 1.47\\
Clean & 1.61 & 1.65 & 1.73 & 1.29 & 1.57 & 1.02 & 1.57 & 1.25 & 1.34\\
\end{tabular}
\end{table}

\begin{table}
\caption{Conductance $G[X]$ in $G_0 \equiv 2e^2/h$ for a $3d$ transition metal occupied divacancy II (TM@DVII) with N$_2$, O$_2$, H$_2$O, CO, NH$_3$, and H$_2$S adsorbed.  The conductance of the clean TM@DVII is provided for comparison.}\label{TM@DVII_G}
\begin{tabular}{l|ccccccccc}
 & Ti & V & Cr & Mn & Fe & Co & Ni & Cu & Zn\\\hline
N$_2$ & 1.65 & 1.77 & 1.65 & 1.27 & 1.12 & 1.33 & 0.51 & 1.06 & 1.49\\
O$_2$ & 1.65 & 1.29 & 1.47 & 1.54 & 1.15 & 1.49 & 1.00 & 1.04 & 1.34\\
H$_2$O & 1.57 & 1.67 & 1.69 & 1.20 & 1.21 & 1.45 & 1.29 & 1.02 & 1.56\\
CO & 1.68 & 1.79 & 1.52 & 1.29 & 1.07 & 0.61 & 0.69 & 1.17 & 1.56\\
NH$_3$ & 1.61 & 1.69 & 1.67 & 1.19 & 1.18 & 1.50 & 0.86 & 1.01 & 1.60\\
H$_2$S & 1.62 & 1.71 & 1.65 & 1.23 & 1.15 & 1.43 & 0.79 & 1.06 & 1.58\\
Clean & 1.60 & 1.78 & 1.71 & 1.22 & 1.06 & 1.41 & 1.71 & 1.04 & 1.49\\
\end{tabular}
\end{table}

\begin{figure}[!h]
\centering
\includegraphics*[width=\linewidth]{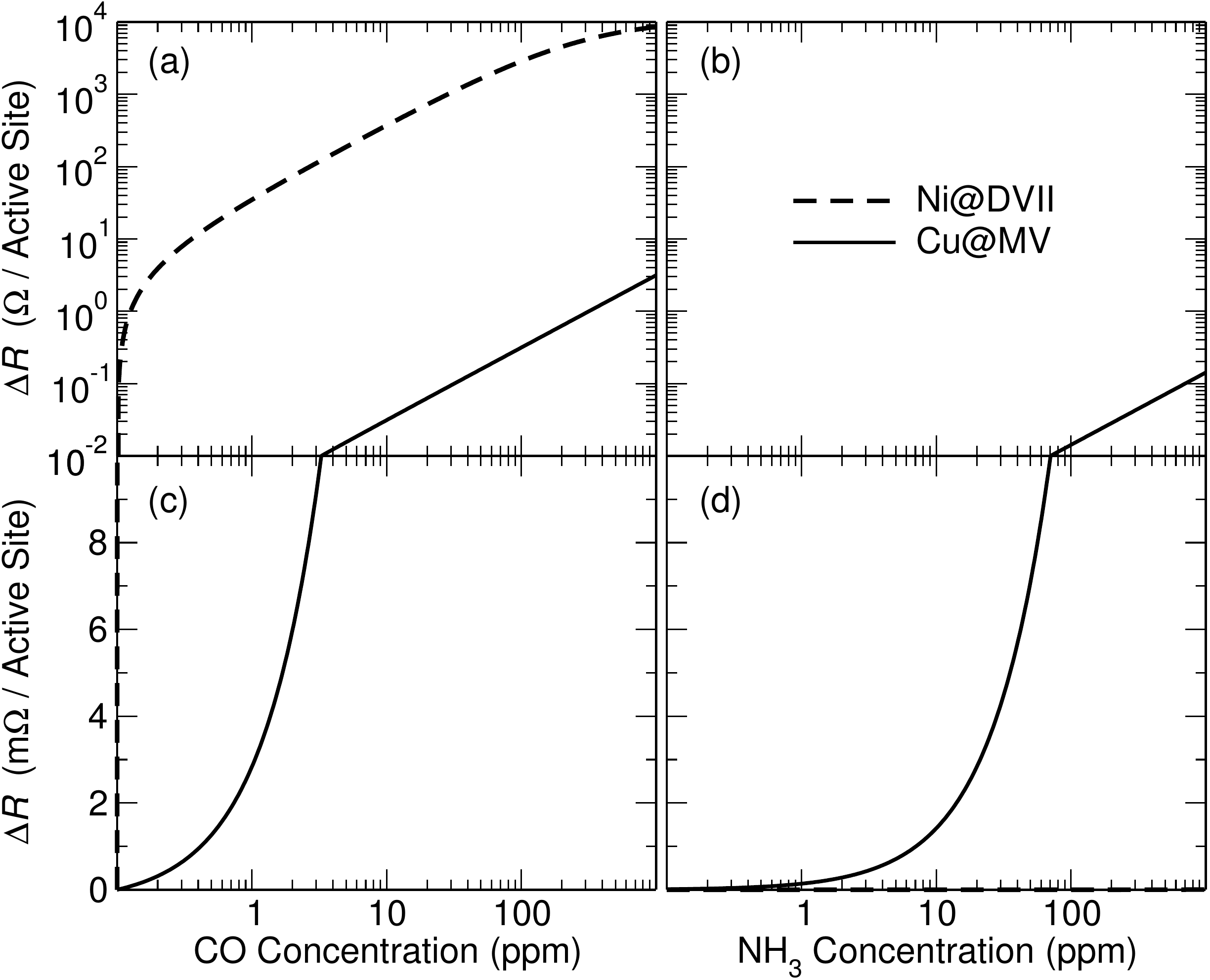}
\caption{%
Change in resistance $\Delta R$ in (a,b) ${\Omega}$ and (c,d) m${\Omega}$ per active site versus (a,c) CO and (b,d) NH$_{{3}}$ concentration in ppm for Ni in a divacancy II (Ni@DVII,{\bf{--~--~--~--}}) and Cu in a monovacancy (Cu@MV,{\bf{---------}}) of a (6,6) carbon nanotube.  The reference concentrations of CO and NH$_{{3}}$ are 0.1 and 0.01 ppm, respectively, in a background of air at room temperature and 1 bar of pressure.}
\label{Fig5}
\end{figure}

We now estimate the resistance of a SWNT containing several impurities
(a specific metal dopant with different molecular adsorbates).  Under
the assumption that the electron phase-coherence length, $\ell_\phi$, is
smaller than the average distance between the dopants, $d$, we may
neglect quantum interference and obtain the total resistance by adding
the scattering resistances due to each impurity separately. The
scattering resistance due to a single impurity is then given by
\begin{equation}
R_s[X]=\frac{1}{G[X]}-\frac{1}{2G_0}.
\end{equation}
Here $G[X] = T[X\mathrm{+TM@VC}](\varepsilon_{\mathrm{F}})$ is the Landauer conductance of the pristine SWNT with a
single metal dopant occupied by molecule $X$, i.e. the transmission at the Fermi level $\varepsilon_{\mathrm{F}}$, $1/(2G_0)$ is the
contact resistance of a (6,6) SWNT, and  $G_0 \equiv {2} e^{{2}}/h$ is the quantum of conductance.  The calculated conductances are provided in Tables \ref{TM@MV_G}, \ref{TM@DVI_G}, and \ref{TM@DVII_G} for a molecule $X$ adsorbed on a $3d$ TM occupied monovacancy, divacancy I, and divacancy II, respectively.

We may now obtain the average change in resistance $\Delta R$ of an active site as a function of target molecule concentration.  As discussed in Ref.~\cite{Juanma}, this change in resistance is reasonably well described by
\begin{equation}\label{eq.totalres}
\Delta R \approx \sum_{X} R_s[X](\Theta[X,C] - \Theta[X,C_0]), 
\end{equation}
where $C$ is the concentration, and $C_0$ is the concentration at standard temperature and pressure, as given in Table \ref{Table1}.

In Figure \ref{Fig5} we show $\Delta R$ for a single active site (Ni or Cu) as a function of target molecule concentration (CO or NH$_{{3}}$).  Keeping in mind that a change in concentration from 1 ppm to 50 ppm amounts to a change from allowable to toxic concentrations for both CO and NH$_{{3}}$, we see that this sensor design should work effectively for both target molecules.  

\section{Conclusions}

In summary, we have presented a general model for the design of chemical nanosensors which takes the adsorption energies of the relevant
chemical species and their individual scattering resistances as the
only input. On the basis of this model we have performed a
computational screening of TM doped SWNTs, and demonstrated using \emph{ab initio} calculations how a combined Cu and Ni doped metallic SWNT device may work effectively as a multifunctional sensor for both CO and NH$_{{3}}$.  Further, we find that by varying the metal dopant, we obtain ``another handle'' for tuning the sensitivity of our sensor.  The methodology employed may also be applied to other nanosensor designs and different environments, and demonstrates the potential of computational screening studies for the design of chemical nanosensors \emph{in silico}.

\section*{Acknowledgments}
We acknowledge funding through the European Research Council Advanced
 Grant DYNamo (ERC-2010-AdG -Proposal No. 267374), Spanish ``Juan
de la Cierva'' and Jos{\'{e}} Castillejo programs (JCI-2010-08156), Spanish Ministerio de Ciencia e Innovac{\'{\i}}on (FIS2010-21282-C02-01,
FIS2010-19609-C02-01), Spanish ``Grupos Consolidados UPV/EHU del
Gobierno Vasco'' (IT-319-07, IT-366-07), and ACI-Promociona
(ACI2009-1036), ARINA, NABIIT and the Danish Center for Scientific
Computing. The Center for Atomic-scale Materials Design (CAMD) is
sponsored by the Lundbeck Foundation. 




\end{document}